\documentclass[aps,prb,twocolumn,superscriptaddress]{revtex4-1}
\usepackage{bm}
\usepackage{graphicx}
\usepackage{graphics}
\usepackage{amsmath}
\usepackage{amsfonts}
\usepackage{amssymb}
\usepackage{makeidx}
\usepackage[colorlinks=true,citecolor=blue,linkcolor=blue,linktocpage=true,pagebackref=false]{hyperref}
\hypersetup{colorlinks=true,citecolor=blue,linkcolor=blue,filecolor=blue,urlcolor=blue}
\usepackage{color,soul} 
\usepackage{float}
\graphicspath{{figures/}} 
\begin{document}

\newcommand{\be}   {\begin{equation}}
\newcommand{\ee}   {\end{equation}}
\newcommand{\ba}   {\begin{eqnarray}}
\newcommand{\ea}   {\end{eqnarray}}
\newcommand{\ve}  {\varepsilon}
\newcommand{\Dis} {\mbox{\scriptsize dis}}

\newcommand{\state} {\mbox{\scriptsize state}}
\newcommand{\band} {\mbox{\scriptsize band}}

%%%%%%%%%%%%%%%%%%%%%%%%%%%%%%%%%%%%%%%%%%%%%%%%%%%%%%
\title{Spin relaxation in disordered graphene: \\
Interplay between puddles and defect-induced magnetism}

\author{V. G. Miranda}

\affiliation{Instituto de F\'{\i}sica, Universidade Federal
  Fluminense, 24210-346 Niter\'oi, RJ, Brazil}
\author{E. R. Mucciolo}

\affiliation{Department of Physics, University of Central Florida,
  Orlando, FL 32816-2385, USA}

\author{C. H. Lewenkopf}

\affiliation{Instituto de F\'{\i}sica, Universidade Federal
  Fluminense, 24210-346 Niter\'oi, RJ, Brazil}

\date{\today}

\begin{abstract}
We study the spin relaxation in graphene due to magnetic moments
induced by defects. We propose and employ in our studies a microscopic
model that describes magnetic impurity scattering processes mediated
by charge puddles. This model incorporates the spin texture related to
the defect-induced state.  We calibrate our model parameters using
experimentally-inferred values. The results we obtain for the spin
relaxation times are in very good agreement with experimental
findings. Our study leads to a comprehensive explanation for the short spin
relaxation times reported in the experimental literature. We also
propose a new interpretation for the puzzling experimental observation
of enhanced spin relaxation times in hydrogenated graphene samples
in terms 
% of disorder configurations that lead to an enhanced 
% coupling of the magnetic moments.
of a combined effect due to disorder configurations that lead to an 
increased coupling to the magnetic moments and the tunability of 
the defect-induced $\pi$-like magnetism in graphene.
\end{abstract}

\maketitle

%%%%%%%%%%%%%%%%%%%%%%%%%%%%%%%%%%%%%%
\section{Introduction}
\label{sec:introduction}
%%%%%%%%%%%%%%%%%%%%%%%%%%%%%%%%%%%%%%

The spin properties of graphene \cite{Han2014,Roche2014review,
  Pesinreview, Soriano2015} are one of the many fascinating aspects of
this material. The combination of carbon's low atomic number and the
planar structure of graphene's lattice entails a very low intrinsic
spin-orbit coupling, in the range of 1 $\mu$eV \cite{Han2014,
  Roche2014review, Pesinreview}, and negligible hyperfine interactions. 
  As a consequence, pioneering
theoretical studies predicted that electrons could travel for long
distances and long times in graphene without spin flipping
\cite{Han2014, Roche2014review, Pesinreview}. This property would make
graphene a very interesting candidate for passive spintronics devices
\cite{Han2014, reviewFabian2007, reviewFabian2004}. However, since the
first successful experimental demonstration of injection of a
spin-polarized current in graphene \cite{Tombros2007}, the measured
spin relaxation times have been found to be orders of magnitude
smaller than the theoretical predictions
\cite{Han2014,Roche2014review,Soriano2015}. The small relaxation times
observed in Ref. \onlinecite{Tombros2007} were confirmed by other
experiments with different setups \cite{Wojtaszek2013, Maassen2013,
  Omar2015, Barbaros2013, KawakamiPRL2011, KawakamiPRL2012,
  Drogelernlet2016}. A large theoretical effort has been devoted to
unravel this puzzle \cite{ErtlerPRB2009, OchoaPRL2012, FedorovPRL2013,
  DVTuanNATPHYS2014, BundesmannPRB2015, CazalillaPRB2016,
  KochanPRL2014, Soriano2015, EversPRB2015, Harju2017,
  DVTuanscirep2016, CummingsPRL2016, PedersenPRB2015, DVTuanPRL2016}
and different mechanisms have been proposed for the enhanced spin
relaxation, such as charge puddles \cite{ErtlerPRB2009,
  DVTuanscirep2016, Harju2017}, impurity-induced enhancement of spin
orbit interaction \cite{FedorovPRL2013, BundesmannPRB2015,
  DVTuanNATPHYS2014, CazalillaPRB2016} and magnetic impurities
\cite{KochanPRL2014, Soriano2015, EversPRB2015, Harju2017}.

Light adatoms such as hydrogen and lattice defects (vacancies) have
been proposed to give rise to $\pi-$like magnetism in graphene
\cite{YazyevPRB2007,Yazyev2010}. The origin of this kind of magnetism
relies on the bipartite character of the lattice
\cite{Yazyev2010,Miranda16} and on electron-electron interactions
\cite{Lieb89}. Whenever an imbalance between sublattices $A$ and $B$
occurs $N=|N_{A}-N_{B}|\neq 0$ zero modes appear
\cite{Pereira08,Abrahams1994}, and upon the presence of
electron-electron interactions the ground state of the system is
magnetic with total spin $S=N/2$ \cite{Yazyev2010,Lieb89}. For the
case of a single impurity, it has been shown theoretically that the
magnetism is associated with the wave function of the zero mode, which
extends over many lattice sites \cite{Yazyev2010, HaasPRB2011,
  Nanda12, PalaciosRossierBrey, KochanPRL2014, Miranda16}.  This
magnetism is unusual since, ordinarily, magnetism is related to atoms
with $d$ and $f$ unfilled shells and is very localized around such
atoms. $\pi-$like induced magnetism due to hydrogen adatoms
\cite{Gonzalez-Herrero2016} or vacancies \cite{Zhang2016} were
reported recently in scanning tunneling microscopy experiments and the
magnetic texture confirmed \cite{Gonzalez-Herrero2016}.

Kochan and collaborators \cite{KochanPRL2014} proposed that magnetic
impurities in graphene act as hot spots, flipping the spins of the
electrons as they are scattered by these impurities. Small spin
relaxation times have been measured in hydrogenated
\cite{Wojtaszek2013, Barbaros2013, KawakamiPRL2012} samples and in
graphene with vacancies \cite{KawakamiPRL2012}. Although other
mechanisms such as spin-orbit coupling are predicted to arise in
hydrogenated graphene \cite{CastroNetoPRL2009, GmitraPRL2013,
  Barbaros2013} and compete with the magnetic-induced spin flipping
for fast spin relaxation, several studies point to the latter as the
dominant mechanism in these systems \cite{Soriano2015,
  BundesmannPRB2015, EversPRB2015}.

It is difficult to capture the spatial texture of the impurity-induced
magnetism in graphene with an effective impurity Hamiltonian. Previous
studies made use of {\it ab initio} calculations to evaluate the
microscopic structure of the system and extract the parameters of an
effective impurity tight-binding Hamiltonian
\cite{GmitraPRL2013,KochanPRL2014,Soriano2015,PedersenPRB2015}. The
pioneering approach developed by Kochan and collaborators
\cite{KochanPRL2014} indicate that the largest contribution to
spin-flip processes comes from the exchange term at the hydrogen site,
justifying neglecting the contribution coming from farther
sites. Although this simplification allows for an elegant analytical
solution of the problem, it is not clear what are the effects of
disregarding the extended character of the induced magnetism. The
authors of Ref. \onlinecite{KochanPRL2014} needed to perform an ad hoc
convolution to treat the effect of electron-hole puddles when fitting
their theory to the experimental data
\cite{KochanPRL2014}.

A very recent study \cite{Harju2017} has shown that close to the Dirac
point the energy dependence of the spin relaxation time has a strong
reliance on the range of the magnetic impurities. This result points
to the relevance of considering the extended nature of the magnetism
induced by vacancies or hydrogen adatoms in graphene. Recently, two
studies \cite{Soriano2015,Harju2017} have proposed different models to
treat spin-flip process due to magnetic impurities; however, the
spatial range and modulation of the magnetic profile in these models
was chosen arbitrarily.

In this paper we address the issue of spin relaxation in graphene due
to magnetic moments induced by defects. We derive a comprehensive
microscopic formalism that accounts for the interplay between local
chemical potential fluctuations, which give rise to charge puddles,
and magnetic impurities, and hence incorporates the spin texture
related to the defect-induced state. In our model, the local chemical
potential fluctuations not only modify the density of states close to
the charge neutrality point, but also play a key role in the spin flip
scattering mechanism.
%We put forward a comprehensive and systematic
% investigation of the microscopic model parameters to unravel the role
%played by different parameters on the spin relaxation times in
%graphene. 
Our study leads to a novel, alternative explanation for the short spin
relaxation times reported in experimental literature and sheds some
light on the dominant mechanism of spin relaxation in hydrogenated
graphene.

The paper is organized as follows: In Sec.~\ref{sec:theory} we
introduce the tight-binding Hamiltonians we use to model vacancies and
adatoms in graphene in the presence of disorder-induced charge
puddles. Further, we present a microscopic derivation of the Anderson
single impurity model and the spin relaxation times within this
setting. In Sec.~\ref{sec:results} we show the results of our
simulations and analyze how the spin relaxation time $\tau_s$ depends
on the model parameters. The technical details of the numerical
procedure are discussed in the Appendix. We conclude in
Sec. \ref{sec:conclusions} with a discussion of different scenarios to
interpret the experimental results within realistic parameter choices.

%%%%%%%%%%%%%%%%%%%%%%%%%%%%%%%%%%%%%%%%
\section{Theoretical approach}
\label{sec:theory}
%%%%%%%%%%%%%%%%%%%%%%%%%%%%%%%%%%%%%%%%

We study spin relaxation due to two kinds of defects that have been
shown to induce magnetism in graphene, namely, vacancies and hydrogen
adatoms. A single tight-binding model is usually employed to address
both types of defects \cite{YazyevPRB2007,SorianoPRL2011,
  Soriano2015}. Here we treat the two cases separately, analyzing
their similarities and differences.

Taking into account electron-electron interaction and long-range
disorder, we map the model defect tight-binding Hamiltonian onto the
single impurity Anderson model.

%%%%%%%%%%%%%%%%%%%%%%%%%%%%%%%%%%%%%%%%
\subsection{Mapping onto the Anderson Hamiltonian}

To represent a vacancy, we adopt the standard $\pi$-band
nearest-neighbor tight-binding Hamiltonian to describe the low-energy
electronic properties of a graphene sheet,
\begin{equation}
\label{TA1}
H^{\rm vac} = - t \sum_{\langle i, j \rangle} |i\rangle \langle j| + t
\sum_{\langle {\rm v}, i \rangle} | {\rm v}\rangle \langle i| +
\mbox{H.c.},
\end{equation}
where $\langle \cdots \rangle$ restricts the sum over nearest-neighbor
atomic sites and $t=2.8$ eV is the hopping matrix element between
nearest neighbors. Notice that the second term on the right-hand side
of Eq. (\ref{TA1}) removes the defective (v site) from the honeycomb
lattice.

\textit{Ab initio} calculations have suggested that the effect of a
hydrogen adatom bonded on top of a carbon site can be modeled as
\cite{WehlingPRB2009, RobinsonPRL2008, GmitraPRL2013}
\begin{align}
\label{TA1B}
H^{\rm ad} = & - t \sum_{\langle i, j \rangle}\left( |i\rangle \langle
j| + |j\rangle \langle i| \right) \nonumber\\ & + \epsilon_{H}|\rm
ad\rangle\langle {\rm ad}| + T_{\rm ad-C} \left(| {\rm v}\rangle
\langle \rm ad| + | {\rm ad}\rangle \langle \rm v| \right),
\end{align}
where $\epsilon_{H}$ denotes the adatom on-site energy and $T_{\rm
  ad-C}$ is the hopping between the adatom and the ${\rm v}$-th carbon
site to which the adatom binds. There are different estimates for
these parameters \cite{WehlingPRB2009, RobinsonPRL2008,
  GmitraPRL2013}. We take $\epsilon_{H}=0.16$~eV and $T_{\rm
  ad-C}=7.5$~eV, which provide the best fit to the {\it ab
  initio}-calculated band structures \cite{GmitraPRL2013}.

The models represented by Eqs.~\eqref{TA1} and \eqref{TA1B} give rise
to midgap states pinned at (or close to) the Dirac point and localized
around the defective site \cite{Pereira08, Nanda12,
  PalaciosRossierBrey, HaasPRB2011, Yazyev2010, GmitraPRL2013,
  WehlingPRB2009, Abrahams1994}. More explicitly, the eigenvalue
problem associate to these model systems can be cast as
\be
\label{TA2}
H^{\rm def} | \nu\rangle = \ve_\nu | \nu\rangle,
\ee
where the superscript ``def" stands for ``vac" (vacancy) or ``ad"
(adatom), with
\be
\label{eq:TA3}
|\nu\rangle=\left\{ \begin{array}{ll} \lbrace|\phi\rangle\rbrace, &
  \quad \mbox{for} \quad \ve_{\nu} \neq \ve_{\rm m}, \\ |0\rangle, & \quad
  \mbox{for} \quad \ve_{\nu} = \ve_{\rm m}\,.\end{array} \right.  \ee
The energy of the midgap state $\epsilon_{\rm m}=0$ for the vacancy
\cite{Pereira08} and $\epsilon_{\rm m}\approx 10$~meV for the adatom
\cite{GmitraPRL2013}. Notice that the states $\left\{|\phi \rangle
\right\}$ and $|0\rangle$ form a complete basis, where the former are
extended states and the latter is a 'quasilocalized' one
\cite{Pereira08}.

We use this complete eigenbasis to define the projection operators,
\be
P = \sum_{\phi \neq 0} |\phi\rangle\langle \phi | \quad \mbox{and} \quad 
Q = |0\rangle\langle 0 |,
\ee
where $P+Q=1$ by construction. 
%This allows us to decompose the Hamiltonian as $ H^{\rm def} =
% (P+Q)H^{\rm def}(P+Q) \equiv H^{\rm def}_{PP}+H^{\rm
% def}_{PQ}+H^{\rm def}_{QP}+H^{\rm def}_{QQ}, \end{align}
Since the $|0\rangle$ and $|\phi \rangle$ states are orthonormal, the
coupling terms are zero, $H^{\rm def}_{PQ}=H^{\rm def}_{QP}=0$. As a
consequence, in this scenario there is no spin-flip scattering of the
band states by the impurity state. This picture changes if long-range
disorder is included in the model, as it has been studied to explain
the Kondo effect in defective graphene \cite{Miranda14} and Fano
resonances in Lieb-like optical lattices \cite{FlachEPL2014}.

%Kondo effect in this scenario is not possible because due to
%orthogonality there is no coupling between the localized $|0\rangle$
%and extended states $\lbrace |\nu \rangle \rbrace$
%states\cite{Miranda14}. The picture changes if one remember that real
%samples are disordered \cite{Mucciolo10} and includes this ingredient
%in the modelling. Here we include the effect of long range disorder
%through the diagonal term
Long-range disorder is ubiquitous in graphene samples. Its main
sources are puddles, i.e., inhomogeneous real-space fluctuations of
the chemical potential that arise mainly due to charge transfer
between graphene and its substrate or trapped impurities reminiscent
from the sample preparation process \cite{MartinNP2008, XueNMAT2011,
  ZhangNP2009, Mucciolo10}. To account for such an effect we add to
$H^{{\rm def}}$ the onsite terms
\be
\label{TA4}
V_{\rm dis} = \sum_{i} |i \rangle V_{i} \langle i |,
\ee
where $V_i$ is given by \cite{AdamPRB2011,Mucciolo10}
\be 
\label{TA5} 
V_i \equiv V_{\rm dis}(\bm{r}_{i})=  \sum_{j=1}^{N_{\rm
G}} W_{j}\, e^{-\dfrac{(\bm r_i - \bm R_j)^{2}}{2 \xi^{2}}}.
\ee
$V_i$ corresponds to the local potential at the site $i$ due to
$N_{{\rm G}}$ long-range Gaussian potentials randomly placed at
positions $\bm{R}_j$. The parameter $\xi$ denotes the disorder
potential range and $W_j$ follows a Gaussian distribution with
\be 
\label{TA6}
\left\langle W_{i}\right\rangle=0 \quad \mbox{and} \quad \left\langle
W_{i}W_{j}\right\rangle =(\delta W)^{2} \delta_{ij},
\ee
where $\delta W$ defines the disorder strength.

The typical disordered charge puddle patterns observed experimentally
\cite{MartinNP2008, XueNMAT2011} are reproduced by the model defined
above by taking $\xi=10$ nm, $\delta W=56$ meV, and $N_{\rm G}/N_{\rm
  tot}\agt 0.04\%$ for silicon dioxide (SiO$_2$) substrates and
$\xi=30$ nm, $\delta W=5.5$ meV, and $N_{\rm G}/N_{\rm tot}\agt
0.004\%$ for hexagonal boron nitride (hBN) ones
\cite{Burgos2015,DVTuanscirep2016}, where $N_{\rm tot}$ denotes the
number of carbon atoms in the sample.

Employing the projectors algebra for the full Hamiltonian incremented
by the long-range term potential in Eq. \eqref{TA4}, we find
\be
\label{TA7}
H=H^{\rm def}+V_{\rm dis} = 
% (P+Q)H(P+Q) \equiv 
H_{PP}+H_{PQ}+H_{QP}+H_{QQ},
\ee
The terms in Eq. \eqref{TA7} split into three types: a localized
``impurity" contribution,
\be
\label{TA8}
H_{QQ}= QHQ= (\ve_{\rm m}+\ve_{0}^{\prime}) |0 \rangle \langle 0
|\equiv \ve_{r} |0 \rangle \langle 0 |, \ee
where $\ve_{0}^{\prime} = \langle 0 | V_{\rm dis} |0 \rangle$ gives
the energy shift of the midgap state due to puddle disorder; a
coupling contribution,
\be
\label{TA9}
H_{PQ} = \sum_{\phi \neq 0 } |\phi \rangle \langle \phi | V_{\rm dis}
|0 \rangle \langle 0 |
\ee
and a ``band" contribution,
\be
\label{TA10}
H_{PP} = \sum_{\phi \neq 0} |\phi \rangle \ve_\phi \langle \phi | +
\sum_{\phi\neq 0}\sum_{\phi^{\prime} \neq 0} |\phi \rangle \langle
\phi | V_{\rm dis} |\phi^{\prime} \rangle \langle \phi^{\prime}|.  \ee
In general, $\langle \phi | V_{\rm dis} |\phi^{\prime} \rangle\neq 0$,
implying that $H_{PP}$ is not diagonal in the $\{ |\phi \rangle \}$
basis. To map our model onto the single-impurity Anderson model (SIAM), 
we introduce a new $\{ |\beta \rangle \}$ basis where the
'band' term of the SIAM is diagonal. The $\beta$ states are obtained
by diagonalizing $H_{PP}$ in Eq. \eqref{TA10}, namely,
\be
\label{TA11}
H_{P^{\prime}P^{\prime}} | \beta \rangle = \ve_\beta |\beta\rangle,
\ee
where $|\beta\rangle$ and $|\phi\rangle$ are related via $|\beta\rangle =
\sum_{\phi}|\phi\rangle\langle\phi|\beta\rangle$ and the projector
$P^{\prime}$ for the ${|\beta \rangle}$ subspace reads
\be
\label{TA12}
P' = \sum_\beta | \beta \rangle \langle \beta |.
\ee
In this basis the coupling term becomes
\be
\label{TA15}
H_{P'Q} = \sum_\beta |\beta \rangle \langle \beta | V_{\rm dis} |0
\rangle \langle 0| \equiv \sum_\beta |\beta \rangle t_{\beta 0}
\langle 0 |, \ee
where the hopping coefficients $t_{\beta 0}\equiv\langle \beta |
V_{\rm dis} |0 \rangle$ fluctuate with $|\beta \rangle$ and disorder
realization.

{\it Ab initio} calculations suggest more involved tight-binding 
models for vacancies and  adatoms \cite{Nanda12, NandaPRB2016},
that take into account local lattice deformations. We do not
investigate such models here, but the approach we have put 
forward can be easily generalized to incorporate lattice 
reconstruction effects. We do not expect this additional source
of disorder to qualitatively change the results we obtain with
our simple model. 

Equations \eqref{TA8}, \eqref{TA11}, \eqref{TA15} combined give the
single-particle contribution to our SIAM model. The remaining term to
complete the mapping is the interaction contribution, embodied by the
charging energy $U$. A recent work \cite{Miranda16} showed how to
calculate the charging energy of a vacancy-induced state within the
tight-binding formalism and studied $U$ as a function of the system
size. There has been a long debate in the literature
\cite{PalaciosRossierBrey, Palacios2012, Nanda12, NandaPRB2016} on
whether defect-induced $\pi-$like magnetism in graphene due to
vacancies is relevant for realistic samples sizes. The theoretical
results presented in Refs.~\onlinecite{Miranda16,NandaPRB2016} and
recent STM experiments on hydrogenated graphene
\cite{Gonzalez-Herrero2016} and on graphene with vacancies
\cite{Zhang2016} give strong support to the magnetic
scenario. However, there is still some discrepancy between the
theoretical estimates for $U$ \cite{Miranda16} and the values inferred
from experiments \cite{Gonzalez-Herrero2016, Zhang2016}. The origin of
this discrepancy is still not clear. For instance, it has been
speculated that screening effects \cite{AdamPRL2016} due the
environment to which graphene is submitted should be taken into
account for the theoretical modeling to yield estimates closer to the
experiment.

For these reasons, in this work we choose to leave $U$ as a free
parameter and study its influence on $\tau_s$ estimates. We stress
that the typical values used here are consistent with the experimental
scenario \cite{Gonzalez-Herrero2016,Zhang2016}.

%-------------------------------------------------------------------------------------------------
\subsection{Schrieffer-Wolff transformation}
\label{sec:SW}

We write our model in second quantization as
\begin{align}
\label{SW0}
H_{\rm {SIAM}} = & \sum_{\beta,\sigma} \ve^{}_{\beta}\,
c_{\beta\sigma}^{\dagger}c_{\beta\sigma} + \sum_\beta \left( t_{\beta
  0}\, c_{\beta\sigma}^{\dagger}c_{0\sigma} + {\rm H.c.}\right)
\nonumber \\ & + \sum_\sigma \ve_{r}\, c_{0\sigma}^{\dagger}
c_{0\sigma} +
% Un_{0\uparrow}^{\rm dis}n_{0\downarrow}^{\rm dis}.
Un_{0\uparrow} n_{0\downarrow},
\end{align}
where $\sigma = \uparrow,\downarrow$ denotes the spin degree of
freedom. The linear density of states and the bipartite character of
the graphene lattice are incorporated in the model by the extended
states $|\beta\rangle$. The SIAM allows for both charge and spin
fluctuations \cite{Anderson1961,Hewson1993} since the impurity state can
assume the empty, single, or double occupation configurations. Since
we are interested in impurity-induced spin-flip scattering processes,
we focus on the magnetic singly-occupied regime.

We use a Schrieffer and Wolff transformation \cite{SW} to keep only
the magnetic subspace of the model. By doing so we treat the hopping
between the impurity and the band states as a ``perturbation"
\footnote{Strictly speaking this is not a perturbative approach since
  high-order terms contribute to the corrections due to $t_{\beta
    0}$. The approach aims to isolate these interaction terms by
  applying the canonical transformation and eliminating $t_{\beta 0}$
  to first order.} 
and write the SIAM Hamiltonian as
\be
\label{SW1}
H_{\rm SIAM} = H_{0} + H_{1}, 
\ee
where
\be
\label{SW2}
H_{0} = \sum_{\beta \neq 0,\sigma} \ve_{\beta}\,
c_{\beta\sigma}^{\dagger} c_{\beta\sigma}^{} + \sum_\sigma
\ve_{r}\, c_{0\sigma}^{\dagger} c_{0\sigma} + Un_{0\uparrow}
n_{0\downarrow}.
\ee
and
\be
\label{SW3}
H_{1} = \sum_{\beta \neq 0,\sigma} \left( t_{\beta 0}\,
c_{\beta\sigma}^{\dagger} c_{0\sigma} + {\rm H.c.} \right).
\ee
The hopping term can be eliminated up to first order in the coupling
by the similarity transformation \cite{SW}
\be
\label{SW4}
\widetilde{H}_{\rm SIAM} = e^{S}\, H_{{\rm SIAM}}\, e^{-S},
\ee
where $S$ is such that $\left[H_{0}, S\right] = H_{1}$. This
conditions is satisfied for \cite{SW}
\be
\label{SW6}
S=  \!\sum_{\beta \neq 0,\sigma} \!\!
% \!\left\lbrace 
t_{\beta 0}^{}\left[ \frac{n_{0,-\sigma}^{}}{\epsilon_{\beta}^{{}}-\epsilon_{0}^{{}}-U} + 
 \frac{ 1-n_{0,-\sigma}^{}  }{\epsilon_{\beta}^{{}}-\epsilon_{0}^{{}}}  \right]
\!c_{\beta\sigma}^{\dagger}c^{}_{0\sigma} -{\rm H.c}.    
%\right\rbrace. 
\ee

Schrieffer and Wolff restricted their analysis of the resulting
Hamiltonian to the case where a magnetic moment is most probable to
develop in a metallic substrate, i.e, $\epsilon_{0}^{}+U>0$ and
$\epsilon_{0}^{}<0$ \cite{SW}. For graphene, the scenario is much
richer and magnetism can develop also for $\epsilon_{0}^{}>0$
\cite{UchoaPRL2008}.

The low energy effective Hamiltonian now reads,
\be
\label{SW7}
\widetilde{H}_{\rm SIAM} = \widetilde{H}_0+H_{{\rm xc}},
\ee
where $\widetilde{H}_0$ accounts for single-particle terms that result
from the canonical transformation in Eq.~\eqref{SW4} and which can be
incorporated into the Hamiltonian $H_0$ with a suitable energy
renormalization \cite{SW,Phillips2012}. The term $H_{{\rm xc}}$ is the
one we are interested since it is responsible for the spin-flip
scattering. It can be written as \cite{SW}
\be
\label{SW8}
H_{{\rm xc}} = -\sum_{\beta \neq 0 \atop \beta^{\prime} \neq
  0} J_{\beta\beta^{\prime}}
%\Big( 
\sum_{\sigma,\sigma^{\prime}} c_{\beta\sigma}^{\dagger}
{\bm S}_{\sigma\sigma^{\prime}} c_{\beta^{\prime}\sigma^{\prime}}
%\Big)
\cdot 
%\Big( 
\sum_{\sigma^{\prime\prime},\sigma'''}
\!c_{0\sigma^{\prime\prime}}^{\dagger}
{\bm S}_{\sigma^{\prime\prime}\sigma^{\prime\prime\prime}}
c_{0\sigma'''}.
%\Big), 
\ee
Here, ${\bm S}={\bm \sigma}/2$, where the Cartesian components of ${\bm
  \sigma}$ are Pauli matrices, and the exchange coupling reads
\cite{SW}
\begin{align}
\label{SW9}
J_{\beta \beta^{\prime}}= t_{\beta 0}^{} t_{\beta^{\prime} 0}^{\ast} &
\left[ \frac{1}{\ve_{\beta}^{{}}-\ve_{r}^{{}}-U} -
  \frac{1}{\ve_{\beta}^{{}}-\ve_{r}^{{}}}
  \right. \nonumber\\ &\left.+
  \frac{1}{\ve_{\beta^{\prime}}^{{}}-\ve_{r}^{{}}-U} -
  \frac{1}{\ve_{\beta^{\prime}}^{{}}-\ve_{r}^{{}}} \right].
\end{align}
The Hamiltonian in Eq.~\eqref{SW8} is the well-known s-d or Kondo
Hamiltonian \cite{Kondo,Hewson1993}.

%%%%%%%%%%%%%%%%%%%%%%%%%%%%%%%%%%%%%%%%%%
\subsection{Spin relaxation times}

With the spin-flip term $H_{\rm xc}$ in hands, we are now in a
position to obtain the spin relaxation time. By using Fermi's golden
rule to define the transition probability rate to go from a state
$\beta$ and spin $\sigma$ to a state $\beta^{\prime}$ and spin
$\sigma^{\prime}$ due to the magnetic-induced state $|0\rangle$, we
get
\be
\label{SR1}
\mathcal{W}_{\beta,\beta^{\prime}}^{\sigma,\sigma^{\prime};m} =
\frac{2\pi}{\hbar} \left|
T_{\beta,\beta^{\prime}}^{\sigma,\sigma^{\prime};m} \right|^{2}
\delta(\ve_{\beta}^{m}-\ve_{\beta^{\prime}}^{m}), \ee
where the transition probabilities read
\be
\label{SR2}
T_{\beta,\beta^{\prime}}^{\sigma,\sigma^{\prime};m } = \langle
\beta^{\prime},\sigma^{\prime}|H_{\rm xc}^{m}|\beta,\sigma \rangle.
\ee
The superindex $m$ has been added in Eqs. \eqref{SR1} and \eqref{SR2}
to denote that these terms are specific to a given disorder
realization $m$.

Taking $H_{\rm xc}$ from Eq. \eqref{SW8} and inserting into
Eq. \eqref{SR2} one gets
\ba
\label{SR3}
T_{\beta,\beta^{\prime}}^{\sigma,\sigma^{\prime};m } & = &
\sum_{\beta^{\prime\prime}\neq 0} \sum_{\beta^{\prime\prime\prime}\neq
  0} J^{m}_{\beta^{\prime\prime},\beta^{\prime\prime\prime}} {\bm
  S}_{0} \nonumber \\ & & \cdot \sum_{\sigma_{1},\sigma_{2}} \langle
\beta^{\prime},\sigma^{\prime}|
c_{\beta_{\prime\prime,\sigma_{1}}}^{\dagger}{\bm
  S}_{\sigma_{1}\sigma_{2}}c_{\beta_{\prime\prime\prime,\sigma_{2}}}
|\beta,\sigma \rangle \nonumber \\ & = &
J^{m}_{\beta,\beta^{\prime}} {\bm S}_{0}
\cdot {\bm S}_{\sigma\sigma^{\prime}}, \ea
where the operator ${\bm S}_0 \equiv
\sum_{\sigma^{\prime\prime}\sigma^{\prime\prime\prime}}c_{0,\sigma^{\prime\prime}}^{\dagger} {\bm
  S}_{\sigma^{\prime\prime}\sigma^{\prime\prime\prime}}
c_{0,\sigma^{\prime\prime\prime}}$ accounts for
the spin operator acting on the impurity spin.
%As the total spin of the system is conserved, the spin-flip process
%above implies that changing the

Within this formalism, the relaxation rate for a given state
$|\beta\rangle$ is defined as
\be
\label{SR4}
\frac{1}{\tau^{m}_{\sigma,\sigma^{\prime}};\beta} =
\sum_{\beta^{\prime}\neq 0}
\mathcal{W}_{\beta,\beta^{\prime}}^{\sigma,\sigma^{\prime};m }, \ee
and by inserting Eqs. (\ref{SR2}) and (\ref{SR3}) into
Eq. (\ref{SR4}), we obtain
\ba
\label{SR5}
\frac{1}{\tau^{m}_{\sigma,\sigma^{\prime}};\beta} =
\frac{2\pi}{\hbar}|{\bm S}_{0} \cdot {\bm
  S}_{\sigma\sigma^{\prime}}|^{2} \sum_{\beta^{\prime}\neq 0}
|J^{m}_{\beta,\beta^{\prime}}|^{2}
\delta(\ve^{{}m}_{\beta}-\ve^{{}m}_{\beta^{\prime}} ).
\ea

Since the total spin of the system is conserved, the spin prefactor in
Eq. \eqref{SR5} can be straightforwardly computed:
\be
\label{SR6}
|{\bm S}_{0} \cdot {\bm S}_{\uparrow\uparrow} |^{2} = |{\bm S}_{0}
\cdot {\bm S}_{\downarrow\downarrow}|^{2} = |{\bm S}_{0z}|^{2} = 1/4,
\ee
\be
\label{SR7}
|{\bm S}_{0} \cdot {\bm S}_{\uparrow\downarrow}|^{2} = |{\bm
  S}_{0x}-i{\bm S_{0y}} |^{2} = 1/2,
\ee
and
\be
\label{SR8}
|{\bm S}_{0} \cdot {\bm S}_{\downarrow\uparrow}|^{2} = |{\bm
  S}_{0x}+i{\bm S_{0x}} |^{2} = 1/2.
\ee

Spin flipping is associated to the terms $1/\tau_{\downarrow\uparrow}$
and $1/\tau_{\uparrow\downarrow}$ and the spin relaxation rate reads
$1/\tau_{s} = 1/\tau_{\downarrow\uparrow} +
1/\tau_{\uparrow\downarrow}$. The spin-conserving term leads to the
momentum relaxation rate $1/\tau_{p} = 1/\tau_{\uparrow\uparrow} +
1/\tau_{\downarrow\downarrow}$. In view of
Eqs. (\ref{SR5})--(\ref{SR8}), the spin flipping rates reads
\ba
\label{SR8b}
\frac{1}{\tau^{m}_{s;\beta}} = \frac{2\pi}{\hbar}
\sum_{\beta^{\prime}\neq 0}
|J^{m}_{\beta,\beta^{\prime}}|^{2}\delta(\ve^{m}_{\beta} -
\ve^{m}_{\beta^{\prime}} ).
\ea

Since we are only interested in the spin relaxation rate at a given
energy $\ve=\ve_F$, we can define
\ba
\label{SR9}
\frac{1}{\tau^{m}_{s}\left(\ve\right)} & = &
\frac{\sum_{\beta^{\prime}\neq 0} \frac{1}
  {\tau^{m}_{s;\beta^{\prime}}} \delta(\ve
  - \ve^{m}_{\beta^{\prime}})} {\sum_{\beta^{\prime}\neq 0}
  \delta(\ve - \ve^{m}_{\beta^{\prime}})} \nonumber \\ & =
& \frac{1} {\varrho(\ve) }
\sum_{\beta^{\prime}\neq 0} \frac{1}
    {\tau^{m}_{s;\beta^{\prime}}} \delta(
    \ve - \ve^{m}_{\beta^{\prime}}),
\ea
where $\varrho(\ve)$ is the density of states of the disordered graphene
($\beta$-space states)
at the reference energy $\varepsilon$.  Averaging over disorder realizations yields
\be
\label{SR10}
\frac{1}{\tau_{s}\left(\ve\right)} =
\left\langle
\frac{1}{\tau^{m}_{s}\left(\ve\right)}
\right\rangle = \frac{1}{\cal N} \sum_{m^{\prime}}
\frac{1}{\tau^{m}_{s}\left(\ve\right)},
\ee
where ${\cal N}$ is the number of disorder realizations.

%%%%%%%%%%%%%%%%%%%%%%%%%%%%%%%%%%%%%%%%%%
\section{Results}
\label{sec:results}
%%%%%%%%%%%%%%%%%%%%%%%%%%%%%%%%%%%%%%%%%%

To model bulk effects, we consider disordered graphene rectangular
superlattices with periodic boundary conditions. The supercells have
dimensions of $N \times M \gg 1$, where $M$ is the number of sites
along the zigzag crystallographic direction and $N$ is the number of
zigzag chains in the cell. We consider only ${\bm k}=0$ states, since
for large enough supercells any point is representative of the first
Brillouin zone \cite{Zunger1978}. The defect (vacancy or adatom) is
placed in the central lattice site.

The use of periodic boundary conditions implies in a ``selective"
dilution of defects in a single sublattice. It has been found
\cite{Pereira08} that the creation of vacancies with an unbalanced
sublattice concentration affects the peak at $\ve=0$, causes a
broadening of the Van Hove singularities and a gap opening, with the
latter scaling as $\sim x^{1/2}$, where $x$ is the vacancy
concentration.
%
%\VLADIMIR{Return after a more careful look in the literature.}
%\CAIO{This observation can upset the ab initio people, but it is fair. 
%I would rather keep the text as it is.}
%
This finding is in line with the large gaps (of the order of 1eV)
obtained by {\it ab initio} band structure calculations of
hydrogenated graphene \cite{Soriano2015, GmitraPRL2013, KochanPRL2014,
  PedersenPRB2015}. Due to the computation cost, the supercells used
in those calculations are small ($\sim 100$ sites), with corresponding
dilution concentrations of $x\sim 1\%$, hence, generating large gap
energies. These sizable gaps are neither observed in spin relaxation
measurements in graphene with vacancies nor in hydrogenated graphene
\cite{Wojtaszek2013, KawakamiPRL2012, Barbaros2013}, where the typical
defect concentrations are $0.01 -
0.2\%$\cite{Wojtaszek2013,KawakamiPRL2012,Nair2012}.

To overcome this problem, we consider systems with $N_{\rm tot} = 256
\times 256$ sites, corresponding to a defect concentration of $\sim
15$ ppm, in line with previous theoretical studies
\cite{PedersenPRB2015,Soriano2015}. This concentration is about one
order of magnitude smaller than the estimated concentration in
experiments where defects are deliberately introduced in graphene
\cite{Wojtaszek2013,KawakamiPRL2012,Nair2012}. We choose to use a
smaller concentration because we expect the defects concentrations to
be considerably lower in samples where defects are not deliberately
created. We further justify our choice by recalling that there is a
degree of uncertainty in the estimation of the defects concentrations
\cite{JustPRB2014,Nair2012}, and that in real experiments the defects
tend to distribute over the two sublattices and to form other
structures, such as divacancies, that induce defect states around
$\ve=0$ (see, for instance, Ref.~\onlinecite{Pereira08}). Our
supercell size choice leads to an energy gap $\sim 10$ meV which is in
very good agreement with the one observed in a recent experiment (see,
Fig.~2 in Ref.~\onlinecite{KochanPRL2014}).

The system sizes we use also ensure that interaction effects between
different impurities are negligible and the system is in the
paramagnetic limit of isolated magnetic defects, as found
experimentally \cite{Nair2012} and pointed out theoretically
\cite{SorianoPRL2011, PalaciosRossierBrey, YazyevPRB2007}. The results
presented in this section correspond to averages over $10^3$
realizations of puddle disorder.

%%%%%%%%%%%%%%%%%%%%%%%%%%%%%%%%%%%%%%%%%%
\subsection{Vacancies}
\label{sec:vacancies}

\begin{figure}[h!]
\begin{center}
\includegraphics[width=0.75\linewidth]{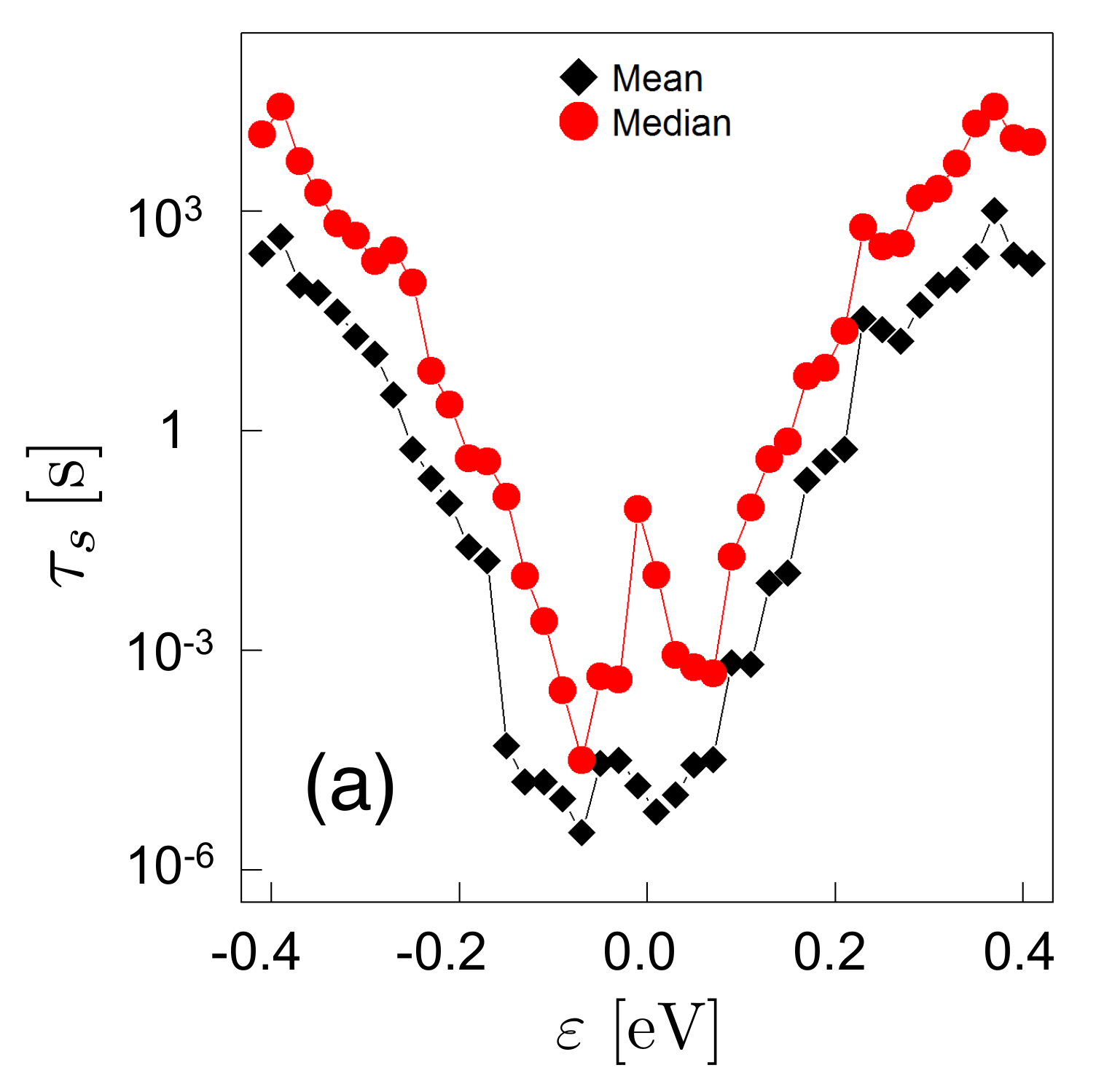}
\includegraphics[width=0.75\linewidth]{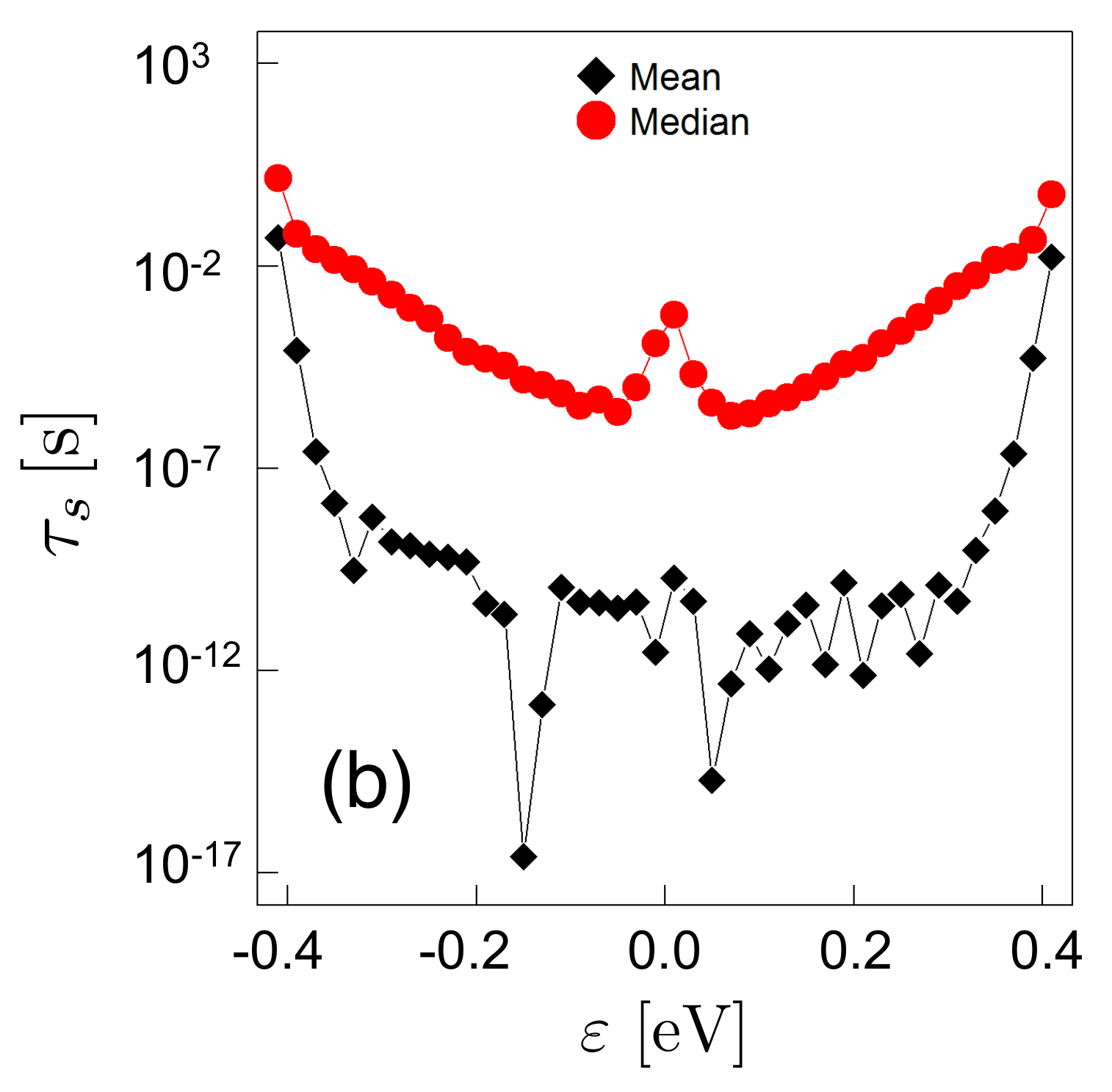}
\vskip-0.2cm
\caption{Spin relaxation time as a function of the Fermi energy for
  hBN (a) and SiO$_2$ (b) substrates. Black diamonds are
  mean-generated values while red circles are median-generated.}
\label{fig:FR1}
\end{center}
\end{figure}

Figure \ref{fig:FR1} shows the spin relaxation times $\tau_s$ due to
vacancy-induced magnetism as a function of the Fermi energy $\ve$ for
graphene on SiO$_2$ and on hBN. We find that the mean values are strongly
influenced by extreme values (outliers). We check this statement by
computing $\tau_s$ using also the median, which is a central-tendency
measure that is robust against outliers. 

%-----------------------------------  F I G U R E  1 ----------------------------------------
%-----------------------------------------------------------------------------------------------

Figure \ref{fig:FR1} also shows that $\tau_s$ (obtained by the mean
value) is relatively low at energies close to the charge neutrality
point or low doping and increases by orders of magnitude at large
doping values. This behavior is similar for both the mean and median
curves, showing that it is not dictated by outliers. 

%\VLADIMIR{The fluctuation due the outliers certainly increases with $\delta W$ but the orders of magnitude variation looks similar for both cases. Maybe to avoid confusion the following sentence should be omitted. Pleas evaluate.} 
% We note that this
% effect becomes more prominent as the disorder strength increases, as
% can be seen in a comparison between the results for hBN
% (Fig.~\ref{fig:FR1}a) and SiO$_2$ (Fig.~\ref{fig:FR1}b).

\begin{figure}[h!]
\begin{center}
\includegraphics[width=0.75\linewidth]{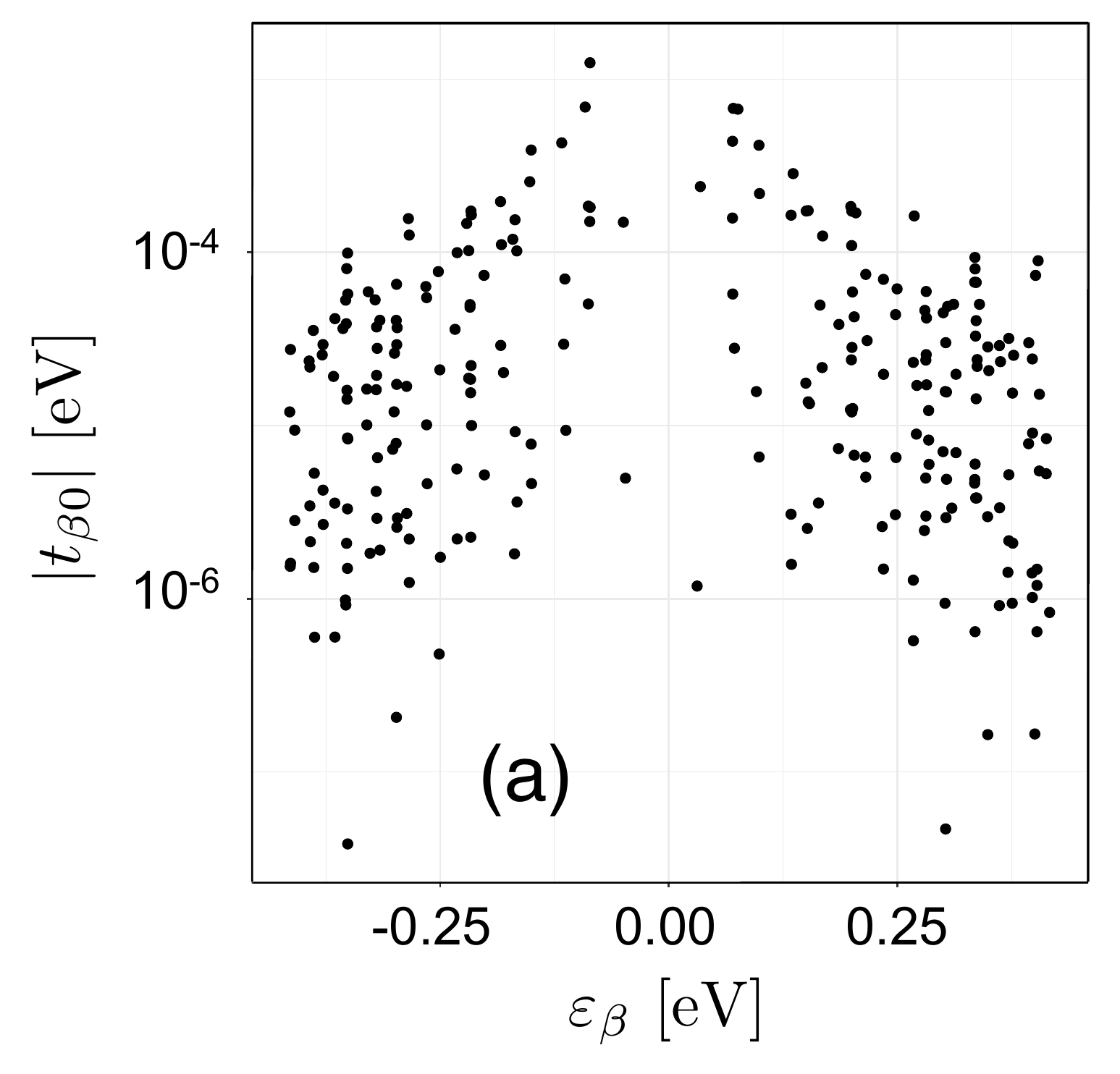}
\includegraphics[width=0.75\linewidth]{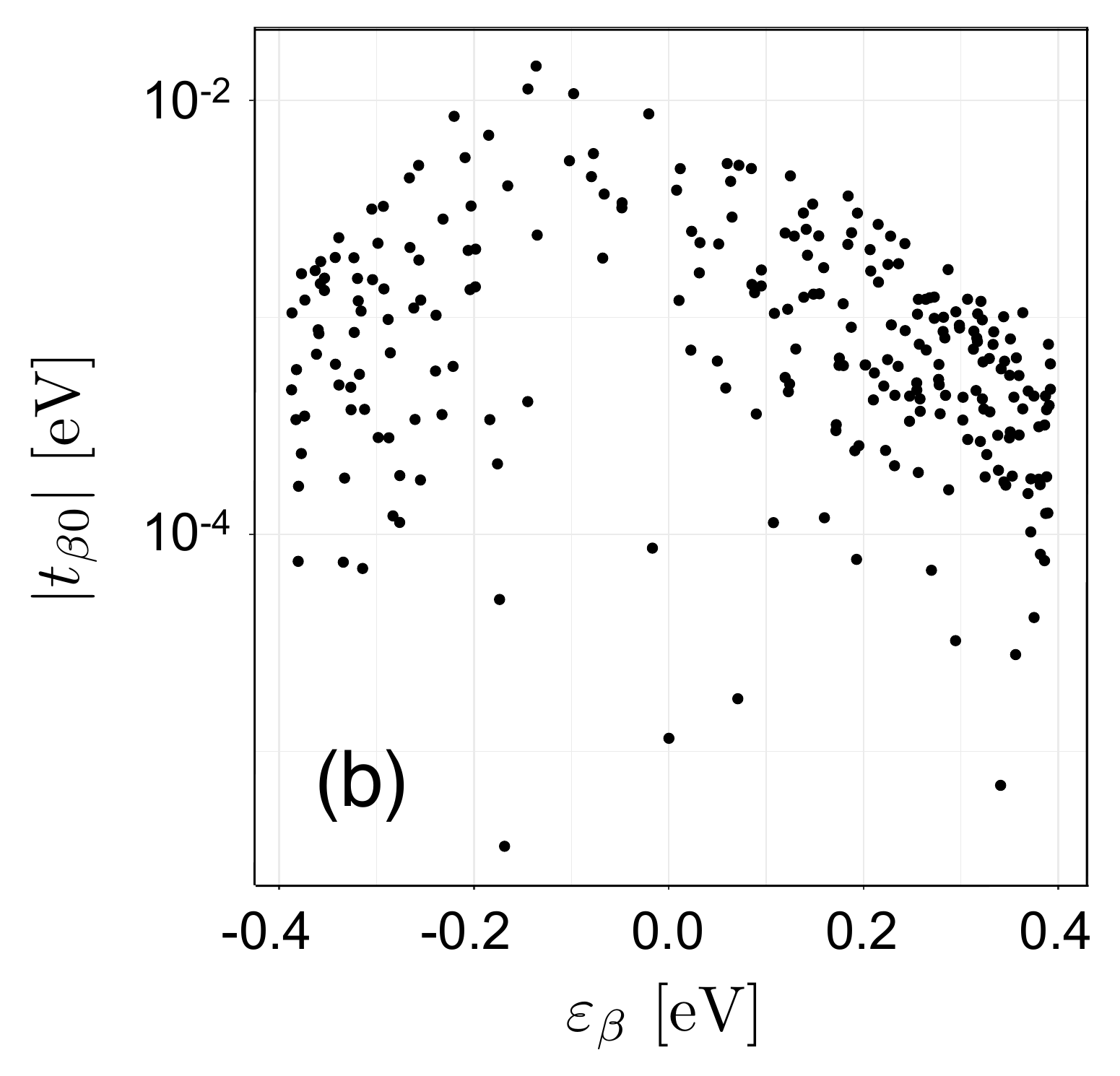}
\vskip-0.2cm
\caption{Energy dependence of $|t_{\beta0}|$ for a single disorder
  realization of the local potential fluctuations due to hBN (a) and
  SiO$_2$ (b) substrates. The magnitude $|t_{\beta0}|$ is
  characterized by large fluctuations and an average exponential decay
  with increasing $|\ve |$.
\label{fig:coupling-no-cutoff} }
\end{center}
\end{figure}

In order to shed some light onto the origin of the behavior seen for
$\tau_s$ in Fig. \ref{fig:FR1}, we study the coupling matrix elements
$t_{\beta 0}$, which is a key element for the calculation of $J_{\beta
  \beta'}$. Figure \ref{fig:coupling-no-cutoff} shows $|t_{\beta0}|$
for a representative single-disorder realization. The coupling matrix
elements $|t_{\beta0}|$ show strong fluctuations and an exponential
average decrease for large $|\ve|$. The {coupling magnitudes} for
 hBN, Fig. \ref{fig:coupling-no-cutoff}a, are smaller than that for
SiO$_2$, Fig. \ref{fig:coupling-no-cutoff}b, which is in agreement
with the behavior of $\tau_s$ in Fig. \ref{fig:FR1}.

%---------------------------------------------- F I G U R E   2 ---------------------------------
%------------------------------------------------------------------------------------------------

We note that magnetic textures observed in STM experiments
\cite{Gonzalez-Herrero2016, Zhang2016, Andrei2016} in defective
graphene extend up to a radius of about 2 nm around the defect and,
hence, are much more localized than the defect-induced state
$|0\rangle$ we obtain in our model. This observation motivates us to
introduce a cutoff radius $R$ to the localized wave function
\footnote{For the sites outside a radius $R$ around the defect
  position, we set the value of the $\left.|0\right>$-state amplitudes
  to zero. The amplitudes of the remaining sites are reweighed such
  that $\left< 0|0\right>=1$.}
and study $\tau_s$ as a function of $R$.

\begin{figure}[h!]
\begin{center}
\includegraphics[width=0.75\linewidth]{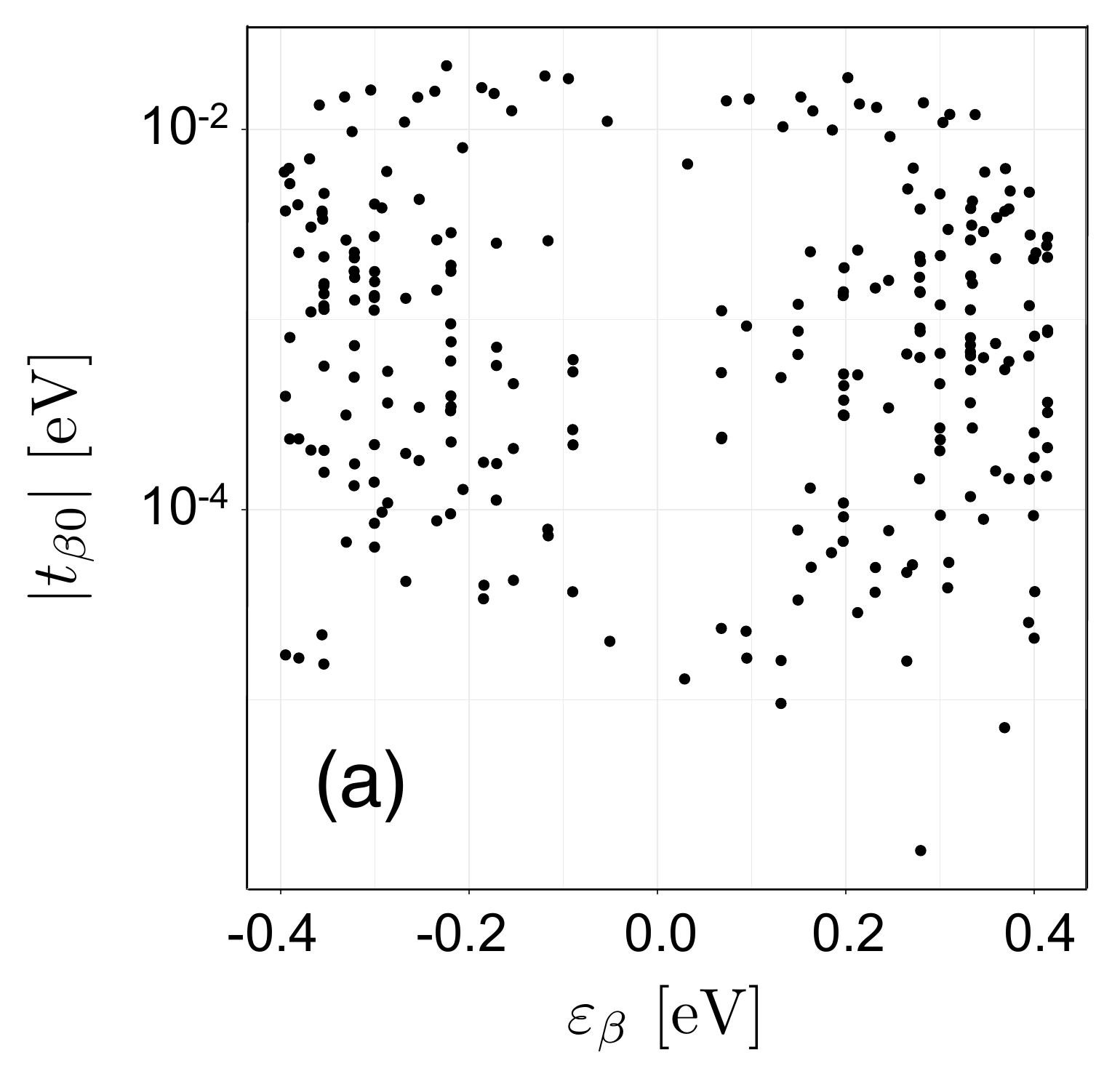}
\includegraphics[width=0.75\linewidth]{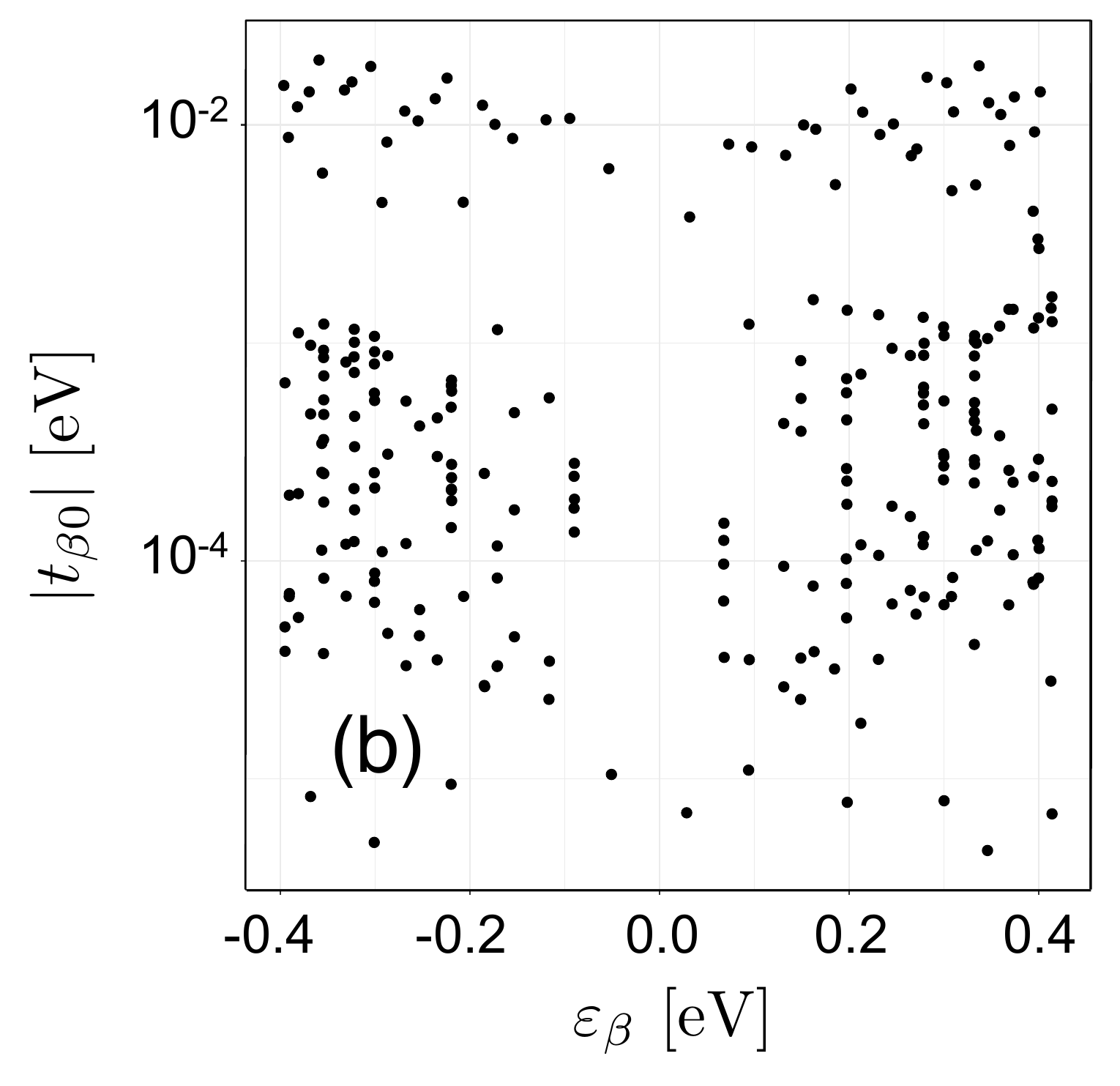}
\vskip-0.2cm
\caption{Energy dependence of $|t_{\beta0}|$ for a single disorder
  realization of hBN considering the impurity state
  confined within a radius (a) $R=1$~\AA ~and (b) $R=2$ nm around
  the defective site. Correlations are quenched and only a random
  pattern is seen. The same change is obtained for the SiO$_2$ case
  (not shown).}
\label{fig:coupling-with-cutoff}

\end{center}
\end{figure}

In Fig.~\ref{fig:coupling-with-cutoff} we plot the results obtained
when $R\approx 1$~\AA, {\it i.e.}, when the wave function is roughly
restricted to the nearest neighbors of the defect and $R\approx 2$ nm,
which is the radius determined in experiments
\cite{Gonzalez-Herrero2016,Zhang2016, Andrei2016}. Remarkably,
Fig. \ref{fig:coupling-with-cutoff} shows that the exponential energy
dependence of the hoppings $|t_{\beta0}|$ vanishes and only a random
pattern is left.

%---------------------------------------------- F I G U R E   3 ---------------------------------
%------------------------------------------------------------------------------------------------

\begin{figure}[h!]
\begin{center}
\includegraphics[width=0.8\linewidth]{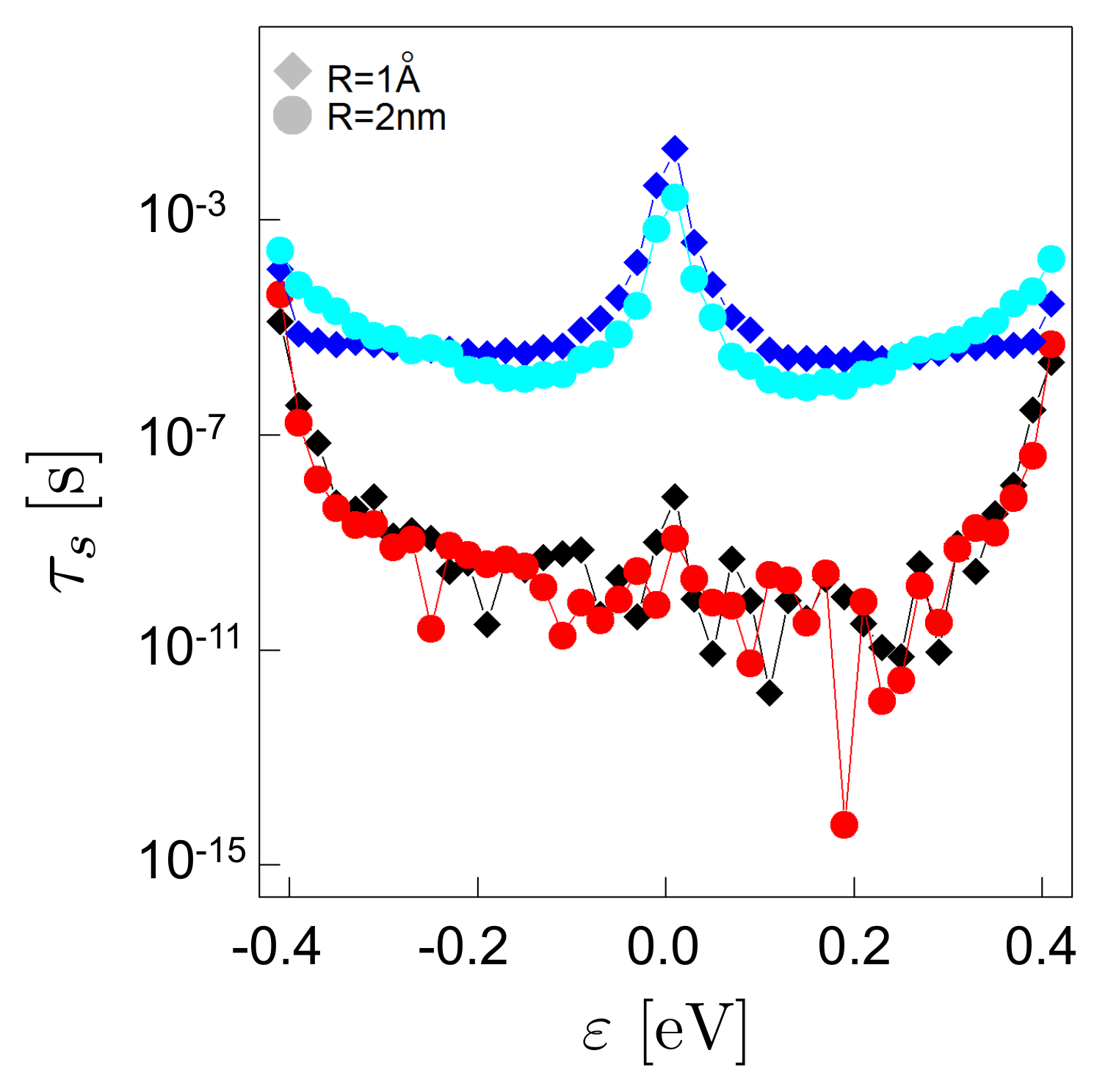}
\vskip-0.2cm
\caption{Relaxation times $\tau_s$ for graphene on SiO$_2$ obtained by
  considering a radial cutoff in the impurity wave functions
  $|0\rangle_{R}$, for $R =1$ \AA ~(diamonds) and $R=2$ nm (circles).
  Black- and red-colored symbols denote mean-generated values while
  blue- and cyan-colored symbols denote median-generated.
\label{fig:tau_vs_R}
  }
\end{center}
\end{figure}

The influence of the extension of the quasilocalized state
$|0\rangle_{R}$ on $\tau_s$ is shown in Fig. \ref{fig:tau_vs_R} for
$R=1$ \AA~ and $R=2$~nm for graphene on SiO$_2$. The values of
$\tau_s$ at low doping ($\ve$ close to the Dirac point) is similar to
those of the extended midgap state $|0\rangle$, Fig.~\ref{fig:FR1}.
In contrast, the exponential dependence of $\tau_s$ observed for
larger values of $|\ve|$ in the $|0\rangle$ case is strongly
suppressed when considering $|0\rangle_R$-states. This is in line with
the above discussion of the coupling matrix $t_{\beta0}$. We note that
the dependence on $R$ is also weak. A comparison between the curves
for the mean and median shows that some outliers still push the mean
estimates down compared to the median estimates and leads to larger
fluctuations. These results are in very good agreement with
experimental measurements
\cite{Barbaros2013,Wojtaszek2013,KawakamiPRL2012}.

For hBN, our simulations (not shown here), when using $\delta W=5.5$
meV, give $\tau_s$ values larger (by a factor 10 to 100 times) than
those reported in the literature
\cite{Barbaros2013,KawakamiPRL2012,Wojtaszek2013,Drogelernlet2016}.
This observation suggests that the effect of outliers is less influent
for cleaner samples. However, experiments show that the disorder
strength $\delta W$ can increase as one approaches the charge
neutrality point \cite{AdamPRL2016} and hence a more detailed
experimental characterization of the puddles strength is necessary
before ruling out the spin relaxation due to magnetic impurities in
hBN.

%---------------------------------------------- F I G U R E   4 --------------------------------
%------------------------------------------------------------------------------------------------

The peak structure, such as the median $\tau_s$ values for $\ve\approx
0$ in Fig. \ref{fig:tau_vs_R}, has been reported in previous
theoretical studies \cite{KochanPRL2014,Soriano2015,PedersenPRB2015}.
In Refs. \onlinecite{KochanPRL2014,Soriano2015}, this peak is
attributed to an enhanced spin-flip resonant scattering process due
the presence of two resonance states, singlet and triplet, with
opposite energies close to $\ve=0$, combined with graphene's vanishing
density of states as $\ve\rightarrow 0$. Since our model deals with
the magnetic regime of the Anderson model, we interpret such behavior
as simply due to the lack of conduction electron states around
$\ve=0$. Also, we show that long-ranged disorder does not quench the
peak, unless the outliers are present in the system, as seen by
comparing the median and mean value estimates for $\tau_s$ in
Fig. \ref{fig:tau_vs_R}.

\begin{figure}[h!]
\label{epsign}
\begin{center}
\includegraphics[width=0.8\linewidth]{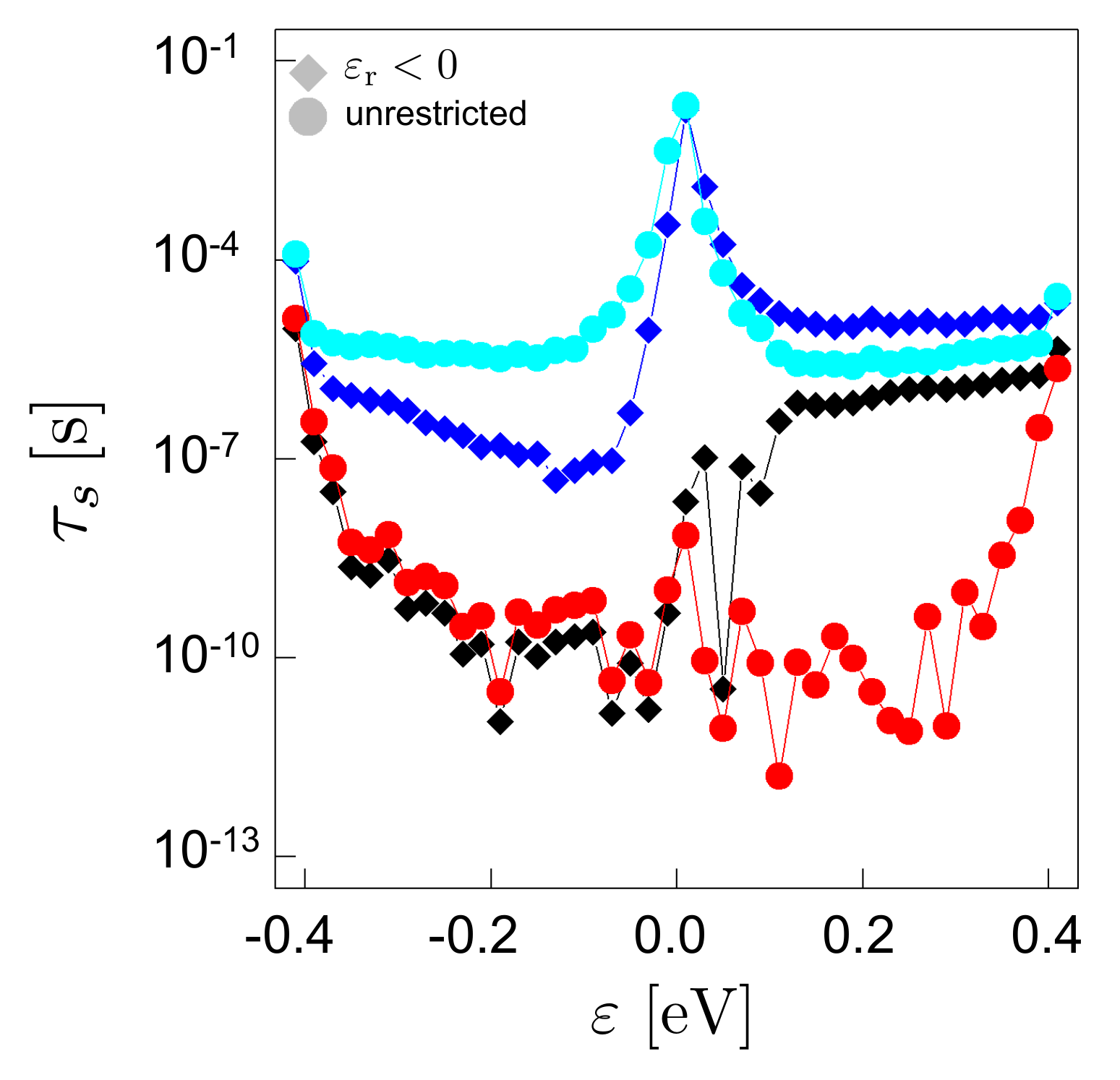}
\vskip-0.2cm
\caption{Comparison between $\tau_s$ for $\epsilon_{\rm r}<0$ only
  and the unrestricted ($\epsilon_{\rm r}\lessgtr0$) case. Black- and
  red-colored symbols denote mean-generated values while blue- and
  cyan-colored symbols denote median-generated. 
\label{fig:tau_s-cutoff}
  }
\end{center}
\end{figure}

In Fig. \ref{fig:tau_s-cutoff}, we study the influence of the midgap
states energies in the calculation of $\tau_s$. As we discussed
earlier, in graphene a localized magnetic impurity remains magnetic
even when $\ve_{r}>0$ \cite{UchoaPRL2008} and the presence of such
magnetic states have a great impact on the $\tau_s$ estimates being
responsible for the approximately particle-hole symmetric behavior of
$\tau_s$ seen in Fig. \ref{fig:tau_s-cutoff} for the unrestricted
curve. The simulations imposing $\ve_{r}<0$ show a huge particle-hole
asymmetry, namely, the values of $\tau_s$ for $\ve>0$ are orders of
magnitude larger than those calculated in the unrestricted case. Such
findings show that an unbalance between $\ve_{r}>0$ and $\ve_{r}<0$
states can be the origin of the asymmetric gate voltage dependence of
$\tau_s$ observed in a recent experiment \cite{Drogelernlet2016}.
Such electron-hole asymmetry in spin relaxation rates has also been
reported for fluorinated graphene \cite{BundesmannPRB2015} where the
resonance energy lies at $-0.2$ eV, in agreement with our discussion.

%---------------------------------------------- F I G U R E   5 --------------------------------
%------------------------------------------------------------------------------------------------

\begin{figure}[h!]
\begin{center}
\includegraphics[width=0.8\linewidth]{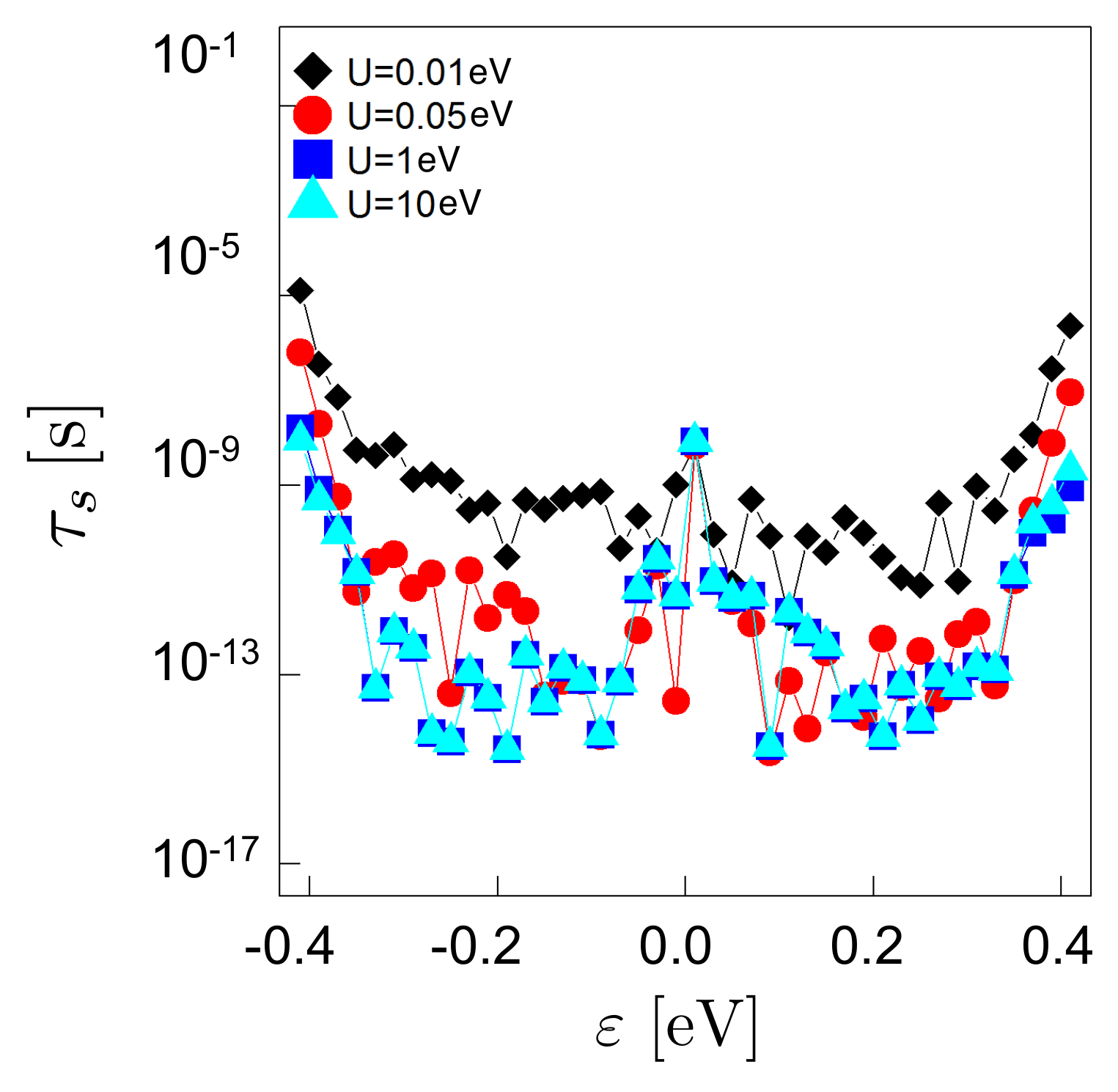}
\vskip-0.2cm
\caption{The dependence of $\tau_s$ on the charging energy $U$. Mean
  value estimates obtained for the $|0\rangle_{R}$ impurity state for
  $R=1$ \AA~in graphene on a SiO$_2$ substrate. Data presented for $U
  = 0.01$ eV (diamonds), $U = 0.05$ eV (circles), $U = 1$ eV
  (squares), and $U = 10$ eV (triangles).
\label{fig:U}
}
\end{center}
\end{figure}

In the simulations shown so far, we have used $U=0.01$eV, a value
inferred from experiments
Ref.~\onlinecite{Gonzalez-Herrero2016,Zhang2016}. However, it is
expected that changes in the impurity environment lead to a different
screening of the impurity \cite{LuicanPRL2014}, and hence different
$U$ values. In Fig.~\ref{fig:U} we study the mean value $\tau_s$
dependence on $U$. Since Fig. \ref{fig:tau_s-cutoff} indicates that
$\tau_s$ is weakly dependent on $R$, we use $|0\rangle_R$ with
$R=1$~\AA.

Figure \ref{fig:U} shows that for $0.01<U<0.1$ eV changing $U$ roughly
leads to a rigid shift of the $\tau_s$ curves. This range of $U$
values is consistent with what is reported by STM experiments for
hydrogenated graphene \cite{Gonzalez-Herrero2016} and graphene with
vacancies \cite{Zhang2016}. Interestingly, increasing $U$ implies in
faster spin relaxation times, showing that screening is an important
ingredient in the computation of the spin relaxation, a feature that
has not been explored in previous studies
\cite{KochanPRL2014,Soriano2015,PedersenPRB2015,Harju2017}. We also
find that the dependence of $\tau_s$ on $U$ becomes negligible for $U
\agt 1$~eV,

%--------------------------------------------- F I G U R E   6 -----------------------------------
%-------------------------------------------------------------------------------------------------

%%%%%%%%%%%%%%%%%%%%%%%%%%%%%%%%%%%%%%%%%%%%%%%
\subsection{Adatom}
\label{sec:adatom}

\begin{figure}[h!]
\begin{center}
\includegraphics[width=0.8\linewidth]{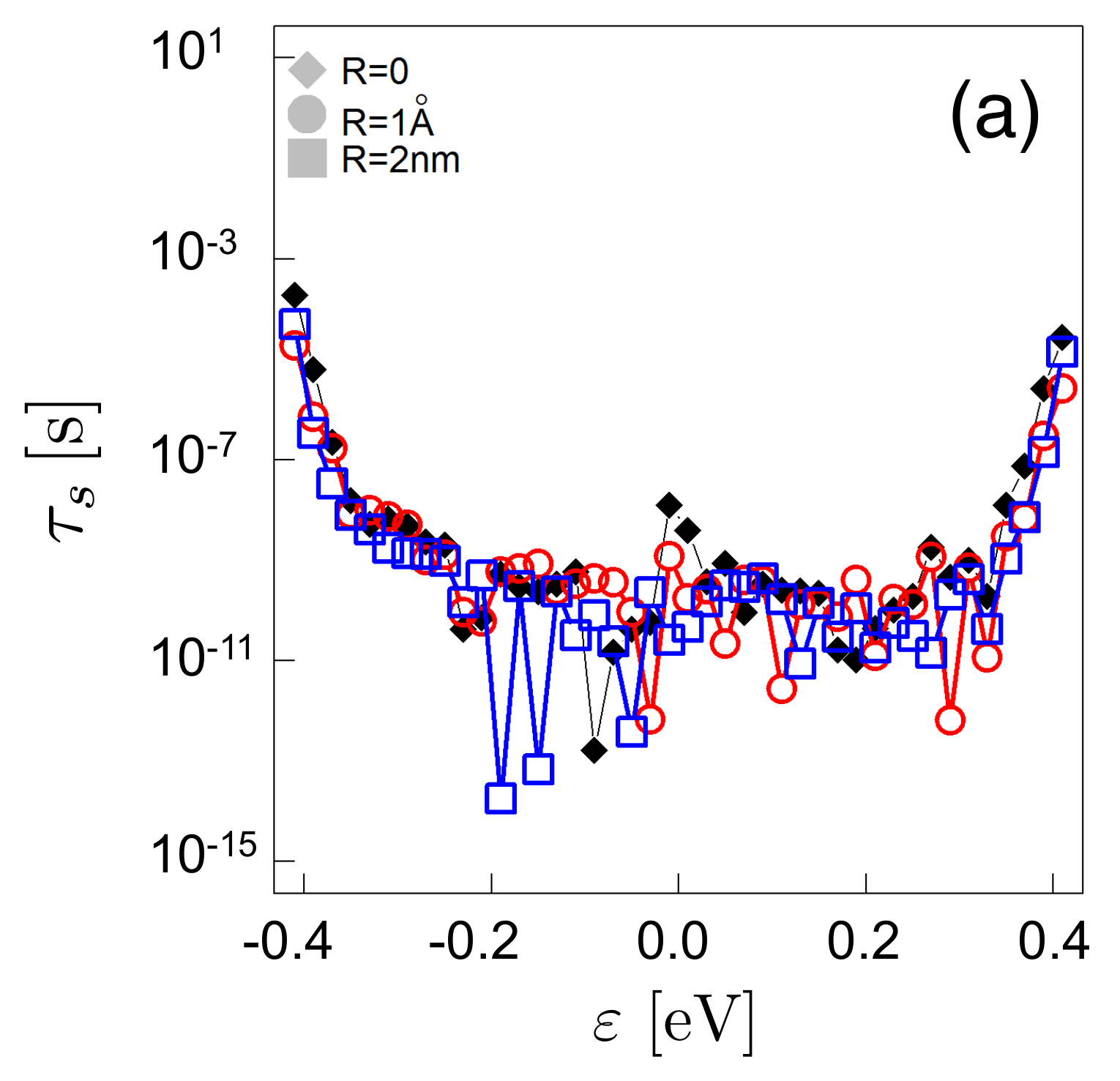}
\includegraphics[width=0.8\linewidth]{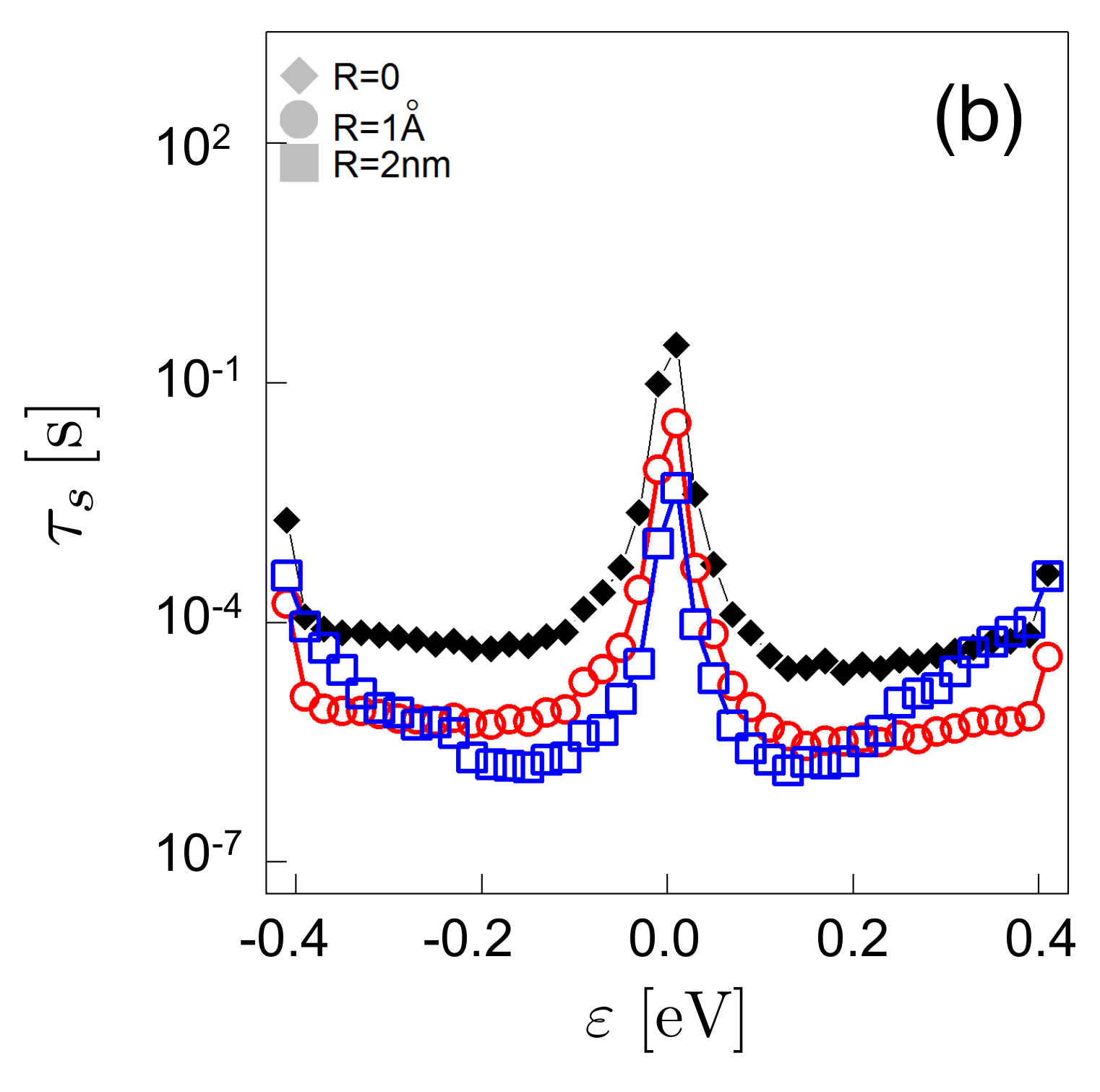}
\vskip-0.2cm
\caption{The $\tau_s$ dependence on energy of different
  impurity-induced states for the adatom model: (a) mean 
  and (b) median $\tau_s$ values for SiO$_2$ substrates. The impurity states
  $|0\rangle_{R}$ considered are $R=0$, $R=1$\AA, and
  $R=2$~nm.}
\label{fig:adatom}
\end{center}
\end{figure}

In this section we analyze the spin relaxation occurring due to the
magnetism induced by hydrogen adatoms in graphene. In
Fig. \ref{fig:adatom}, we plot $\tau_s$ as a function of energy for
different impurity states $|0\rangle_{R}$, with $R=0$, 1\AA, and 2 nm
for SiO$_2$.

The configuration with $R=0$ corresponds to the case where the
magnetism is restricted to the hydrogen adatom site. This case has no
counterpart in the vacancy-induced magnetism discussed in the previous
section. Figure \ref{fig:adatom}a shows that, apart from fluctuations,
the mean $\tau_{s}(\ve)$ is approximately the same for the $R=0$,
$R=1$ \AA, and $R=2$ nm magnetic configurations. On the other hand, a
significant departure between the $R=0$ and the remaining
configurations is observed in the median $\tau_{s}(\ve)$ curves, see
Fig. \ref{fig:adatom}b. For $\ve$ away from the Dirac point the
$\tau_{s}(\ve)$ median values for the $|0\rangle_{R=0}$ state are
roughly an order of magnitude larger than the $R=1$ \AA~and $R=2$ nm
configurations.

Our results indicate that $\tau_s$ is very insensitive to the extent
$R$ of the impurity state, which explains why models with different
kinds of impurity-states find $\tau_{s}(\ve)$ spin relaxation times in
the same range
\cite{KochanPRL2014,Soriano2015,PedersenPRB2015,Harju2017}. Whereas
the median $\tau_{s}(\ve)$ in Fig. \ref{fig:adatom} shows a strong
dependence on $R$, the experimentally observed $\tau_{s}(\ve)$
\cite{Barbaros2013,Wojtaszek2013, KawakamiPRL2012} and the
$\tau_{s}(\ve)$ obtained by different theoretical models
\cite{KochanPRL2014,Soriano2015,PedersenPRB2015,Harju2017} is the one
dictated by the mean and ruled by the ``outliers", which have same
magnitude regardless of the impurity-induced states $R$ for the cases 
where $R<2$ nm, the one reported experimentally \cite{Zhang2016}. Our findings
justifies the use of models with very localized impurity states, as in
Ref. \onlinecite{KochanPRL2014} for the $\tau_s$ calculation, although
the use of such models might become inadequate for the case where the
influence of outliers is not relevant.

%----------------------------------------------- F I G U R E  7 --------------------------------
%-----------------------------------------------------------------------------------------------

%%%%%%%%%%%%%%%%%%%%%%%%%%%%%%%%%%%%%%%%%%%%
\section{Discussion and Conclusions}
\label{sec:conclusions}

In this work we studied the problem of spin relaxation due to
defect-induced magnetism in graphene. In our approach we developed a
microscopic modeling that allows the incorporation of the effect of
charge puddles and its interplay with the defect-induced magnetic
texture. We showed that the puddles play a key role in the spin
relaxation, as it allows for the coupling between the conduction
electrons and the impurity-induced state which is responsible for the
spin flipping scattering process.

Our microscopic derivation consolidates and improves the previous
theoretical approaches \cite{Soriano2015, KochanPRL2014,
  PedersenPRB2015, Harju2017}, since the systematic study of the model
parameters unraveled how the spin relaxation is influenced by
different aspects, such as the quality of the substrate, represented
by the disorder parameters, the extent of the magnetic texture, the
resonances position, and the charging energy $U$. We stress that the
parameters we use in our model are in good agreement with the values
reported in the experimental literature.

\begin{figure}[h!]
\begin{center}
\includegraphics[width=0.8\linewidth]{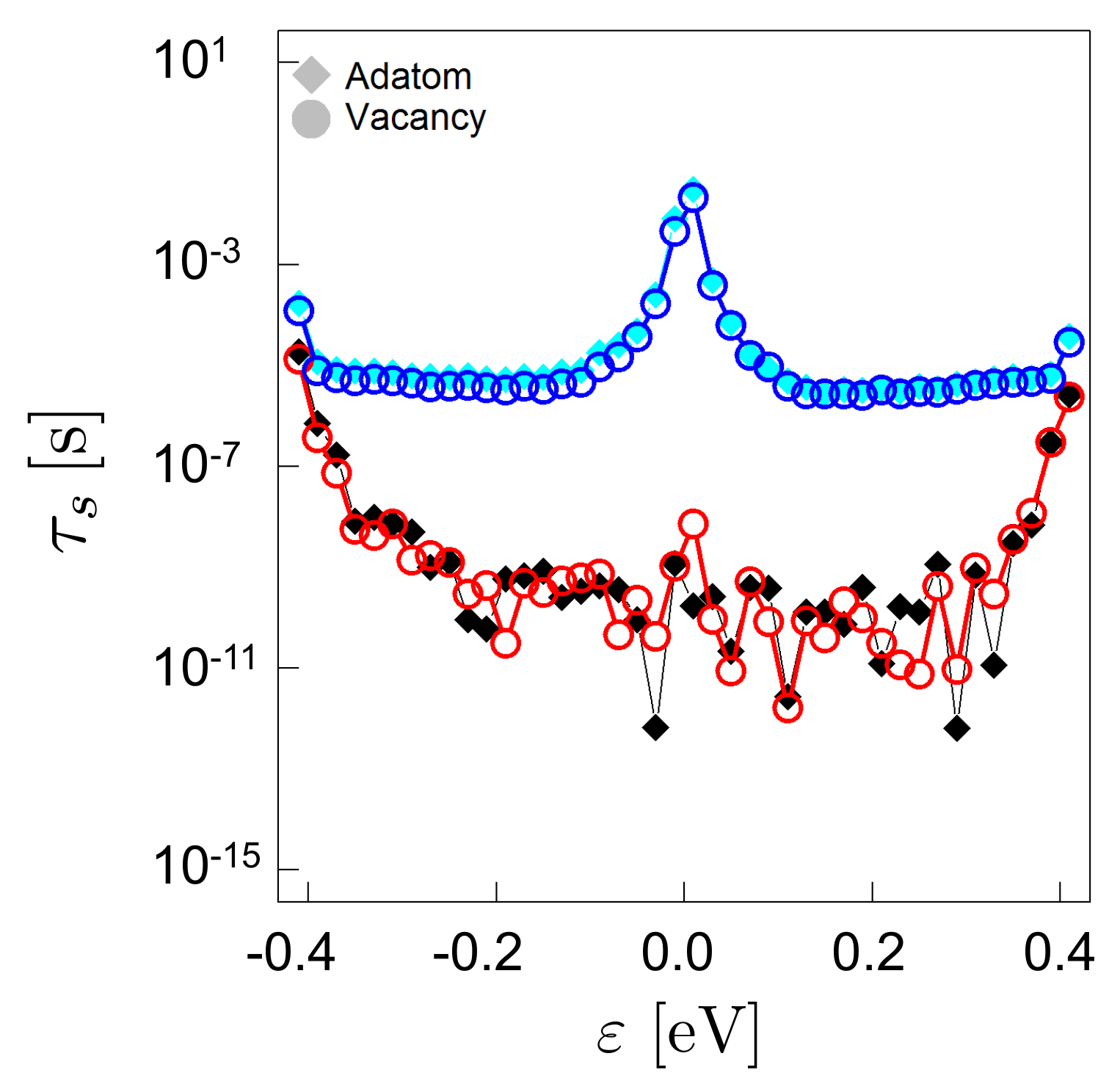}
\vskip-0.2cm
\caption{Spin relaxation time $\tau_s$ as a function of the Fermi
  energy: comparison between the vacancy and the adatom models for
  graphene on SiO$_2$ and $|0\rangle_{R}$, with $R =1$ \AA. }
\label{fig:gran-finale}
\end{center}
\end{figure}

Figure \ref{fig:gran-finale} summarizes our findings, both for vacancy
and adatom-induced spin relaxation in graphene. Using
experimentally-inferred parameter models, we obtained $\tau_{s}$ as a
function of the Fermi energy $\ve$ in the presence of the impurity
state $|0\rangle_{R}$, with $R=1$ \AA, and found that both
spin-relaxation mechanisms show a remarkable agreement. These results
are also in agreement with the experiment reported in
Ref. \onlinecite{KawakamiPRL2012}, which observed a similar spin
relaxation behavior for graphene with adatoms or vacancies.

These results strongly support the picture that spin relaxation in
hydrogenated graphene is dominated by induced magnetism, and not by
induced spin-orbit coupling, contradicting the argument presented in
Ref.~\onlinecite{Barbaros2013}. 
This is in agreement with a recent experiment \cite{Raes2016} that found 
no evidence of a spin-lifetime anisotropy, thus ruling out the spin-orbit coupling 
as a source of ultrafast spin relaxation. 
Our findings also falsify the
conjecture \cite{PedersenPRB2015} that discrepancies in
$\tau_{s}(\ve)$ calculated by different groups could be due to the
defect model choice, namely, adatom
\cite{KochanPRL2014,PedersenPRB2015} or vacancy \cite{Soriano2015}.

%----------------------------------------------- F I G U R E  8 --------------------------------
%-----------------------------------------------------------------------------------------------

Our study also sheds light onto other controversial issues. One of the
main puzzles associated to spin relaxation related to magnetic-induced
defects in graphene is whether the spin relaxation is dominated by the
Elliot-Yaffet (EY) or the Dyakonov-Perel (DP) mechanism
\cite{Soriano2015}. Enhancement of $\tau_{s}(\ve)$ after intentional
introduction of defects in graphene has been reported in
Refs. \onlinecite{Wojtaszek2013,KawakamiPRL2012}, suggesting the DP
mechanism is responsible for the relaxation process. However, the DP
mechanism should be related to random, induced magnetic fields due the
impurities. In this regard, experimental results diverge and a
conflicting scenario about the random magnetic-induced field
interpretation has been questioned
\cite{Wojtaszek2013,KawakamiPRL2012}.

Here, we propose a new alternative explanation for such a
controversy. As has been proposed theoretically \cite{Yazyev2010} and
recently confirmed by STM experiments \cite{Gonzalez-Herrero2016}, the
$\pi$-like magnetism induced by defects in graphene relies on an
unbalance between the number of sites in graphene's sublattices. As a
consequence, such magnetism can be turned on/off by placing defects in
graphene in a unbalanced/balanced manner \cite{Gonzalez-Herrero2016}
among the sublattices.

As we have shown in our study, the main cause of the spin relaxation
is the $\pi$-like magnetism and is essentially independent from the
source of this magnetism, as can be observed in
Fig. \ref{fig:gran-finale} for the case of vacancies and hydrogen
adatoms. We note that other adatom or molecular species have been
proposed to cause a similar $\pi$-like magnetism in graphene, such as
methyl \cite{ZollnerPRB2016} and fluor \cite{Nanda12}. The fact that
different sources may lead to a similar mechanism makes more plausible
that $\pi$-like magnetism is present in different experimental
scenarios, even when the defects are not intentionally created.
 
As has been shown in a recent quantum-interference experiment
\cite{FolkPRL2015}, localized spinfull scatterers are present in
graphene on SiC and are responsible for a fast spin relaxation in such
system. Magnetic-induced spin relaxation has also been observed in
quantum interference experiments in graphene on SiO$_2$
\cite{FolkPRL2013}. Although a detailed characterization of the source
of such magnetic scatterers is still lacking, such similar findings in
different scenarios reinforce our arguments.

We speculate that one or some of the species that can induce
$\pi-$like magnetism in graphene are present in the samples of
Refs. \onlinecite{Wojtaszek2013,KawakamiPRL2012} before the
hydrogenation process and some outliers dominate $\tau_{s}(\ve)$ prior
to the defect exposition. After the hydrogenation, the magnetic
texture due to some of these outliers can be turned off due to the
placement of a hydrogen or vacancy in the opposite sublattice with
respect to the one that contained the magnetic outlier impurity
already in the system. As a result, two spinless scatterers are added,
which can reduce the diffusion in the system but cannot flip the
conduction electrons spin they scatter. Hence, $\tau_{s}(\ve)$
increases (due to the outlier turn off) and the diffusion coefficient
$D$ decreases (due the extra scattering centers after hydrogenation),
in agreement with the behavior reported in
Ref. \onlinecite{Wojtaszek2013}.

This kind of effect seems to be a unique feature of the unusual defect
induced $\pi$-like magnetism in graphene, as shown in a recent study
\cite{Omar2015} of a graphene spin valve device with cobalt porphyrin
(CoPP) molecules. These molecules possess a magnetic moment, which is
extrinsic to graphene. In these systems, no enhancement of
$\tau_{s}(\ve)$ is observed after the introduction of the magnetic
molecules. In fact, the findings in Ref. \onlinecite{Omar2015} enforce
the conjecture that outlying magnetic $\pi$-like scatterers rule the
spin relaxation process since the effect of the introduction of CoPP
on $\tau_{s}(\ve)$ is reported to only be significant in the better
quality samples, and unnoticible for the low-quality ones, which are
the ones with a higher chance for having $\pi-$like spin scatterers be
dominating the relaxation process.

A detailed study of the conditions that lead to the appearance of
outliers ruling $\tau_{s}(\ve)$ is necessary to clarify such
conjectures and will be addressed in a future work.

Since we show that the energy position of the resonances introduced by
defects are essential for the magnitude and symmetry of the
$\tau_{s}(\ve)$ curves, our results also allow us to interpret the
asymmetric gate-voltage dependence of $\tau_{s}(\ve)$ seen in
Ref. \cite{Drogelernlet2016} as an unbalance between resonances with
opposite signs. As different species with different on-site energies
$\ve_0$ can lead to $\pi$-like magnetism
\cite{GmitraPRL2013,ZollnerPRB2016,Miranda16,Nair2012} and the shift
in $\ve_0$ is smaller for systems with lower disorder [see
  Eq. \eqref{TA8}], the asymmetry seen in ultraclean graphene hBN
systems can be due to the predominance of resonances of a given sign.

Finally, our findings suggest an interesting opportunity to control
$\tau_{s}(\ve)$ and hence build spintronic devices via a
functionalization process of graphene with adatoms as the one recently
reported for the precise manipulation of hydrogen adatoms on graphene
via a STM apparatus \cite{Gonzalez-Herrero2016}. By changing the
substrate, the defect species giving rise to the induced magnetism,
and the relative position among the defects in the graphene lattice, a
plethora of possibilities become available for building devices with
different spin properties.

\acknowledgments C.H.L. is supported by CNPq (grant 308801/2015-6) and
FAPERJ (grant E-26/202.917/2015). E.R.M. is supported in part by the
NSF Grant ECCS 1402990.

%%%%%%%%%%%%%%%%%%%%%%%%%%%%%%%%%
\appendix*

%------------------------------------------------------------------------------------------------
\section{Numerical details}

In this Appendix we provide a detailed description of the numerical
analysis used in this work. We focus on two aspects: (i) approximate
diagonalization scheme and (ii) histogram binning.

{\it Approximate diagonalization scheme:}
The computation of $\tau_s$, as outlined in Sec.~\ref{sec:theory},
requires two steps: First, the diagonalization of $H^{\rm def}$, which
gives the $\left\{ |\phi \rangle \right\}$ and $|0\rangle$ basis
states. Second, a basis change from the $\left\{ |\phi\rangle
\right\}$ to the $\left\{ |\beta\rangle \right\}$ subspace. The latter
step requires one to build and diagonalize $H_{PP}$, which is a dense
matrix. As a consequence, considering that disorder averaging is
necessary, the approach becomes prohibitive already for relative small
lattice sizes around $10^3$ sites.

%----------------------------------- F I G U R E  A 1  --------------------------------
\begin{figure}[ht]
\label{FAN1}
\begin{center}
\includegraphics[width=0.75\columnwidth]{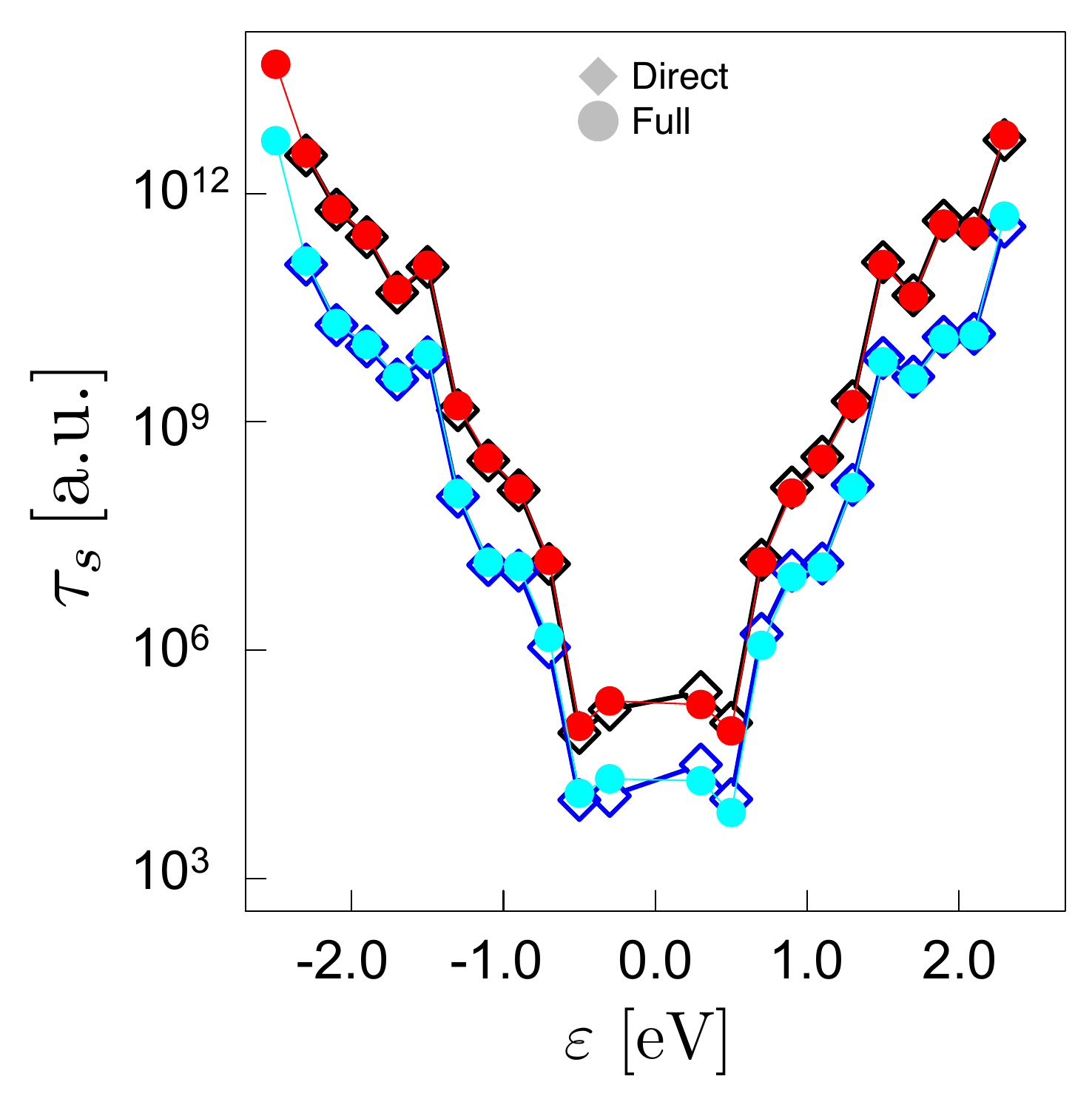}
\vskip-0.2cm
\caption{Comparison between the direct (diamonds) and full (circles)
  diagonalization approaches for a lattice with $N_{\rm tot}=40 \times
  40$ for $\delta W =0.002t$, corresponding to a hBN substrate.
  Estimates are obtained using the median (black and red) and mean
  (blue and cyan).
  \label{fig:comp_methods}}
\end{center}
\end{figure}
%-----------------------------------------------------------------------------------------

To access large system sizes, we introduce a subtle change in the
approach summarized above. We bypass the basis transformation $\left\{
|\phi \rangle \right\} \rightarrow \left\{ |\beta \rangle \right\}$
and directly diagonalize $H^{\rm def}+V_{{\rm dis}}$, which involves
dealing only with sparse matrices. As a result, for every disorder
realization we compute a set of orthogonal extended states
$\{|\tilde{\beta}\rangle\}$ and a quasilocalized state
$|\tilde{0}\rangle$. We replace the $|\tilde{0}\rangle$ state by
$|0\rangle$, the midgap state defined in Eq.~\eqref{eq:TA3}, and take
$\epsilon_{\beta} \rightarrow \epsilon_{\tilde{\beta}}$ and $t_{\beta
  0} \rightarrow t_{\tilde{\beta}0}$. Figure \ref{fig:comp_methods}
shows that the full and the direct diagonalization schemes are in very
good agreement, since $\epsilon_{\beta}\approx
\epsilon_{\tilde{\beta}}$ and $t_{\beta 0}\approx t_{\tilde{\beta}
  0}$.

Notice that the approximation scheme also requires two diagonalization
steps. However, the diagonalization of $H^{\rm def}$ to obtain
$|0\rangle$ is done only once and the diagonalization of $H^{\rm
  def}+V_{\rm dis}$ involves only sparse matrices, which leads to a
huge memory saving and allows for an efficient use of the Lanczos
algorithm to obtain the eigenstates $|\tilde{\beta}\rangle$ in a small
energy window around $\ve=0$, in accordance with experimental values.
\footnote{In the direct diagonalization, we model vacancies by adding
  an on-site energy $V_{\rm v} \gg 1$. The result is equivalent to
  that obtained using Eq.~\eqref{TA1} and the full diagonalization
  scheme, but is more practical for a Lanczos diagonanization.}

%There are a few differences between the modeling of the hydrogen
%adatom and the vacancy: for the vacancy the hoppings between the
%defective sites and its nearest neighbors are removed and an on-site
%energy $V\rightarrow \infty$ is added in the defect. The use of this
%divergent potential showed necessary so that the Lanczos Algorithm
%provided results in conformity with the full diagonalization
%scheme. When modeling the hydrogen adatom impurity the hopping
%elements between the nearest neighbors of the defective carbon site
%are not removed are there is an additional hopping between this
%carbon site and the adatom $T=7.5eV$ taken from DFT models
%\cite{FabianSOC}. The presence of the adatom increases the number of
%sites by $1$ and hence the Hamiltonian dimension is $N_{tot}+1$ and
%this extra element term also carries the onsite energy
%$\epsilon_H=0.16$eV \cite{FabianSOC}. In the diagonalization of
%Hamiltonian $H_0 + U_{\rm dis}$, extra on-site energies are added due
%to the disorder potential calculated according to Eq. \ref{disorder
%pot term}

{\it Histogram approach to calculate $\tau_s$:}
The Fermi golden rule, Eq.~\eqref{SR1}, enforces elastic scattering by
a Dirac delta function in energy. In any practical calculation, the
latter needs to be smoothen. Our strategy is to use a ``histogram
approach": For each disorder realization we construct a histogram of
the energies $\varepsilon_{\beta}$. We label the midpoint of each bin
of the histogram as $\varepsilon_i$. Within each histogram bin we
evaluate the spin relaxation time per eigenstate, according to
Eq. (\ref{SR8b}). Averaging $\tau^m_{s;\beta}$ inside each interval
leads to the $\tau_{s}^m(\ve)$, where the energies are given by the
set of the histogram midpoints.

%----------------------------------- F I G U R E  A 2  --------------------------------
%\begin{figure}[h!]
%\label{FBIN}
%\begin{center}
%\includegraphics[width=0.3\linewidth]{figA2a-spin-relax.pdf}
%\includegraphics[width=0.8\linewidth]{figA2b-spin-relax.pdf}
%\includegraphics[width=0.3\linewidth]{figA2c-spin-relax.pdf}
%\vskip-0.2cm
%\caption{Spin relaxation time as a function of energy for different
%  bin sizes choices.}
%\end{center}
%\end{figure}
%-----------------------------------------------------------------------------------------

The bin size choice is such that it is not large enough to overcome
the gap and small enough to reproduce the low energy density of states
of graphene. In this paper we take bin widths of $20$ meV.

%%\bibliography{spinrelaxrefs}

\end{document}